\documentclass[]{aa}  
\usepackage[utf8]{inputenx}
\usepackage{graphicx}
\usepackage{fixltx2e}
\usepackage{url}
\usepackage{tikz}
\usepackage{grffile}

\newcommand{\calaraltothanks}{Based on observations
    collected at the Centro Astronómico Hispano Alemán (CAHA) at Calar
    Alto, operated jointly by the Max-Planck Institut für Astronomie
    and the Instituto de Astrofísica de Andalucía (CSIC)}

\usepackage{txfonts}
\defcitealias{Christensen2006}{CJW06}
\bibpunct{(}{)}{;}{a}{}{,} 
\begin{document} 
\title{Where is the fuzz? Undetected Lyman $\alpha$ nebulae around QSOs at
  z$\sim$2.3\thanks{\calaraltothanks} } \author{Edmund Christian
  Herenz \inst{1} \and Lutz Wisotzki\inst{1} \and 
  Martin Roth \inst{1} \and
  Friedrich Anders \inst{1} 
          }

   \institute{Leibniz-Institut für Astrophysik Potsdam (AIP) \\
     An der Sternware 16, 14482 Potsdam
              \email{cherenz@aip.de}
             }

   \date{Received \dots; accepted: 17 February 2014}

   \abstract{We observed a small sample of 5 radio-quiet QSOs with 
   integral field spectroscopy to search for possible extended
   emission in the Ly$\alpha$ line. We subtracted the QSO point
   sources using a simple PSF self-calibration technique that
   takes advantage of the simultaneous availability of spatial
   and spectral information. In 4 of the 5 objects we find no significant traces
   of extended Ly$\alpha$ emission beyond the contribution of the QSO
   nuclei itself, while in UM~247 there is evidence for a weak 
   and spatially quite compact excess in the Ly$\alpha$ line at several kpc 
   outside the nucleus. For all objects in our sample we estimated
   detection limits for extended, smoothly distributed Ly$\alpha$ emission
   by adding fake nebulosities into the datacubes and trying to recover them
   after PSF subtraction. Our observations are consistent with other studies
   showing that giant Ly$\alpha$ nebulae such as those found recently 
   around some quasars are very rare. Ly$\alpha$ fuzz around typical 
   radio-quiet QSOs is fainter, less extended and is therefore much 
   harder to detect. The faintness of these structures is consistent
   with the idea that radio-quiet QSOs typically reside in dark matter 
   haloes of modest masses.}


   \maketitle

\section{Introduction}
\label{sec:introduction}

The circum-galactic medium (CGM), defined as gas outside of the main 
stellar body of galaxies but still within the virial radii of their 
dark-matter haloes, is of crucial importance in galaxy evolution.
It may act as a reservoir for fuelling star formation in the galaxy, 
and it is also subject to feedback processes that expell material from the
galaxy. If violent enough, this feedback may in turn heat up the CGM and thus
prevent it from contributing further to the formation of stars. Observing
the CGM at high redshifts will hence provide relevant insights about 
galaxy formation.

One observational approach to study the `cold' gas phase of the
CGM ($T\la 10^4$~K) at high redshift uses absorption signatures
against background sources. This has provided important statistical 
constraints on several properties of the CGM
\citep[e.g.][]{Adelberger2005,Hennawi2006,Steidel2010}. However,
the spatial distribution of the CGM in individual galaxy cannot be
captured with this method.  

An alternative approach is to map the CGM in the Lyman $\alpha$
emission line. Several mechanisms have been identified that should
lead to Ly$\alpha$ emission from the CGM: Cooling of infalling
gravitationally heated gas \citep[e.g.,][]{Haiman2000}, cooling
following superwind-driven shocks \citep[e.g.,][]{Taniguchi2000} and
-- possibly most important for our investigation -- Ly$\alpha$
fluorescence induced by exposure to UV radiation. While the
metagalactic UV background alone is predicted to produce only a very
faint glow that is probably out of reach for the current generation of
optical instruments \citep[e.g.,][]{Kollmeier2010}, Ly$\alpha$
fluorescence caused by the much stronger UV radiation from a QSO
should boost the emission into the detectable regime
\citep{Rees1988,Haiman2001,Kollmeier2010}. Searching for Ly$\alpha$
signatures of the CGM around luminous QSOs is the topic of the present
study.

\cite{Haiman2001} estimated that a $z\sim3$ QSO harbouring a 
$5\times10^{11} \mathrm{M}_\odot$ halo should be surrounded by 
Ly$\alpha$ fuzz extending radially outwards to 2\arcsec{}--3\arcsec{},
at a surface brightness of $\sim 5 \times 10^{-17}$
erg\,s$^{-1}$cm$^{-2}$arcsec$^{-2}$. At least in this model,
Ly$\alpha$ fuzz is predicted as a generic property of high-$z$
QSOs, where the surface brightness of this fuzz depends only on the 
mass of the halo. More recent theoretical work suggests, however, 
that a substantial fraction of gas is accreted within filamentary 
cold streams \citep[][and references therein]{Dekel2009,Faucher-Giguere2011,Rosdahl2012}. 
If these streams are optically thick to ionizing radiation, they will develop a
highly ionized skin in the presence of a quasar. This skin then acts
like a mirror converting up to two thirds of the incident ionizing radiation into
Ly$\alpha$ photons. In this scenario the expected
surface brightness then depends on the ionizing photon flux produced
by the quasar and the projected spatial configuration of the
streams \citep{Kollmeier2010,Hennawi2013}. 

Revealing extended Ly$\alpha$ structures around QSOs requires a proper
subtraction of the PSF-broadened nuclear component, which will
outshine -- even under good seeing conditions -- the expected CGM
signal close to the QSO. Observationally, this makes the detection of 
circum-QSO Ly$\alpha$ fuzz much harder than searching for Ly$\alpha$ `blobs', now
routinely found in large-area narrow-band surveys 
\citep[][]{Steidel2000,Matsuda2004,Saito2006,Ouchi2009,Matsuda2011,Erb2011,Prescott2012,Prescott2013}.
Many of these `blobs' have no obvious central source of ionizing photons
\citep[e.g.,][]{Nilsson2006}, which might make them physically distinct 
from circum-QSO Ly$\alpha$ fuzz. On the other hand, some Ly$\alpha$ blobs may 
also be powered by AGN that are highly obscured along the line of sight
\citep[e.g.][]{Basu-Zych2004,Geach2007,Hayes2011,Martin2014a}; 
such blobs would be physically linked to the fuzz we want to
observe \citep[see also][]{Baek2013}.

As of yet, sizeable samples of extended Ly$\alpha$ emission around
QSOs exist only for radio-loud objects
\citep{Heckman1991a,Heckman1991}, for which such emission seems 
to be a generic property at $z\sim2$--3, with
detections reported up to $z\sim6$ \citep{Roche2014}. However, it
appears likely that a large fraction of extended Ly$\alpha$ emission around radio
loud quasars is driven by interaction with the radio jets. Evidence for this stems
from an observed spatial correlation between radio and Ly$\alpha$
morphology \citep[e.g.,][]{Heckman1991,Humphrey2007} and a
frequent occurrence of similar morphologies in extended \ion{C}{iv}
and \ion{He}{ii} emission \citep{Humphrey2006,Sanchez2009}. 

In order to search for the CGM in QSOs unaffected by radio jets, one
obviously has to resort to radio-quiet objects.  For this class,
however, mostly single object discoveries have been published so far
\citep{Bergeron1999,Bunker2003,Weidinger2004,Weidinger2005,Francis2006,Willott2011,Goto2012,Rauch2013a,Cantalupo2014,Martin2014},
and only one single-object study reported a non-detection at a faint
surface-brightness detection limit \citep{Francis2004}. Very few
programmes aimed at observing (however small) samples
\citep{Christensen2006,Courbin2008,North2012}, and these reported high
detection frequencies of circum-QSO Ly$\alpha$-fuzz:
\cite{Christensen2006} found Ly$\alpha$ fuzz around 4 of 6
  radio-quiet QSOs, while \cite{North2012}, by extending the sample
of \cite{Courbin2008}, found Ly$\alpha$ fuzz around 4 of 6 targets.
So the literature suggests that Ly$\alpha$ fuzz is indeed a rather
generic feature and not a peculiarity around high-redshift radio-quiet
QSOs, even if the number of systematically observed QSOs appears still
too low for certainty.

A major increase in sample size was recently achieved due to the work
of \citet{Hennawi2013}. In a long-slit campaign on 8m-class telescopes
they observed 29 close quasar-quasar pair sightlines, where
Lyman-limit absorption in the background quasar spectrum indicated the
presence of optically thick hydrogen clouds in the CGM of the
foreground $z \sim 2$ radio-quiet quasars. The a priori known presence
of such gas clouds implies that the sample would be positively biased
for detecting `mirrored' Ly$\alpha$ fuzz converted from the quasars
ionizing photons (see above). They find a single quasar surrounded
by a large scale nebula and 10 objects surrounded by small scale
Ly$\alpha$ fuzz.  Surprisingly, none of their spectra showed
Ly$\alpha$ fuzz with the properties expected for optically thick 
`Ly$\alpha$ mirrors'.

For a survey on circum-QSO Ly$\alpha$ fuzz, integral-field
spectroscopy is an optimal observational method since it has
continuous spatial and spectral coverage.  This technique allows for
an optimal subtraction of the PSF-broadened nuclear emission.
Moreover, since spatial and spectral information are obtained
simultaneously in case of a detection, more inferences on the physical
state of the CGM gas can be made. Given the previous success of a
Calar-Alto 3.5m PMAS IFU campaign where 4 extended Ly$\alpha$ nebulae
were detected around 6 radio-quiet QSOs \citep{Christensen2006}
(hereafter CJW06), we initiated a new targeted PMAS IFU campaign to
extend this sample.

The outline of this paper is as follows: In
Sect.~\ref{sec:observations} we describe our PMAS observations and how
we reduced the raw data. In Sect.~\ref{sec:analysis} we explain how we
remove the QSO emission from the datacubes to unveil possible extended
Ly$\alpha$ nebulae. We detail how we estimate surface brightness
upper limits in Sect.~\ref{sec:calc-detect-limits} and 
present the results of those calculations. We summarise our conclusions
in Sect.~\ref{sec:discussion--outlook}.

Conversions of observed to physical quantities always assume
standard $\Lambda$CDM cosmology with $\Omega_\Lambda = 0.7$, $\Omega_M
= 0.3$ and $\mathrm{H}_0 =
70\,\mathrm{km}\,\mathrm{s}^{-1}\mathrm{Mpc}^{-1}$.

\section{Observations \& Data Reduction}
\label{sec:observations}

\begin{table*}
\caption{Observed radio-quiet QSOs: Properties, exposure Times \&
  observing conditions.}
\centering
\begin{tabular}{lccccccccccc}\hline \hline
QSO   & $z_\text{VC}
$ & $z_\text{SDSS}$ & $u_\mathrm{SDSS}$ & $g_\mathrm{SDSS}$&  $t_{\text{exp}}$ [s] & Seeing [\arcsec] & Airmass &  Conditions  \\ \hline
Q0027$+$0103  & 2.310 & 2.313 & 19.27  & 18.50& 7$\times$1800s & 0.9-1.3 & 1.2-1.6 & photometric / grey   \\ 
Q0308$+$0129  & 2.335 &  --  & 19.21  & 18.45 & 7$\times$1800 & 1.0-1.4     & 1.2-1.5 & mainly clear / dark      \\
Q0256$-$0003  & 2.381 & 2.385 & 20.13  & 19.41& 8$\times$1800+1$\times900$ & 0.9-1.1 & 1.3-1.6 & photometric / dark             \\
Q2243$+$0141  & 2.314 &  --  & 18.48  & 17.98 & 8$\times$1800 & 0.9-1.2           & 1.3-1.5 & photometric / grey \\
UM\,247       & 2.333 & 2.333 & 18.85  & 18.10 &5$\times$1800s & 0.9-1.1 & 1.2-1.4 & photometric / dark  \\ \hline \hline
\end{tabular}
\tablefoot{Redshifts from \cite{Veron-Cetty2010}  ($z_\mathrm{VC}$);
and \cite{Paris2012} ($z_\mathrm{SDSS}$). Photometry from \cite{Ahn2012}.}
\label{tab:samp}
\end{table*}

\subsection{Sample selection \& observations}
\label{sec:sample-selection-}

Our sample was selected to be at redshifts around
$z\sim2.3$ so that, in the case of a successful detection of extended
Ly$\alpha$ fuzz, spectroscopic follow-up observations of
rest-frame optical emission lines would be possible. Starting from the
\cite{Veron-Cetty2010} catalog with this constraint, we browsed through
all bright ($m_\mathrm{V}>19$ as in \cite{Veron-Cetty2010}) radio-quiet QSOs at
approximatly the above redshift. In the end we were able to observe 5 
of these targets with the Potsdam Multi-Aperture Spectrophotometer \citep[PMAS;][]{Roth2005} at Calar Alto Observatory during 3 consecutive 
cloudless dark- to grey-time nights in October 2011.  
Table~\ref{tab:samp} lists the redshifts and the SDSS $u$ and $g$ band 
magnitudes of our observed QSOs.

We used PMAS in its 16$\times$16 lens array (LArr) configuration,
giving an 8\arcsec$\times$8\arcsec\ field of view (FoV), (i.e.\ at
0.5\arcsec$\times$0.5\arcsec\ spatial sampling).  Motivated by the
increased sensitivity and larger spectral coverage of the upgraded
PMAS detector \citep{Roth2010}, we mounted the V1200 grating to obtain
the highest possible spectral resolution within the targeted
wavelength range. The CCD was read out unbinned in dispersion
direction.  Exposure times, seeing, airmasses and observing
  conditions of our observations are listed in Table~\ref{tab:samp}.
To ensure the best possible spectral tracing and wavelength
calibration we flanked each on-target exposure by continuum and HgNe
lamp exposures.  At the beginning and end of each night we observed an
\cite{Oke1990} spectrophotometric standard star (BD$+$24d4655 and
G191B2B, respectively). We obtained skyflats during twilight and
several bias frames during the night in idle time (e.g. while
performing acquisitions).

\subsection{Data reduction with \texttt{p3d}}
\label{sec:datar-with-p3d}

To reduce the observations we employed the
\texttt{p3d}-pipeline\footnote{\url{http://p3d.sourceforge.org}}
\citep{Sandin2010,Sandin2012}. We briefly outline in this section how
we applied the different tasks of the pipeline to our data.

For every observing night a masterbias was created with the routine
\texttt{p3d\_cmbias}. The \texttt{p3d\_ldmask} task then produced
dispersion solutions for every arc-lamp frame, fitting
5\textsuperscript{th} order polynomials for the mapping from pixel- to
wavelength space; measured residuals for all arc lines were
$\lesssim 10^{-1}$px.  To determine spectrum traces and
cross-dispersion profiles we applied the \texttt{p3d\_ctrace} method
to every continuum lamp frame.  We created wavelength calibrated
flatfields using \texttt{p3d\_cflatf}. The \texttt{p3d\_cobjex} task
then performed extraction, flat-fielding, and subsequent wavelength
calibration of every target and standard star exposure.  As
detailed by \cite{Sandin2010}, the best signal-to-noise for PMAS LArr
spectra is achieved by using the modified optimal extraction (MOX)
algorithm \citep{Horne1986}.  Due to the large separation of LArr
fiber traces on the CCD, cross-talk correction between fibers is
not needed. \cite{Sandin2012} advises correcting for small spectra
trace shifts between continuum-lamp and on-target exposures for
achieving highest fidelity with MOX extraction, therfore we switched on
median recentering in \texttt{p3d\_cobjex}. Compared to extractions
where we experimentally turned off this feature, we saw a
clear improvement in signal-to-noise, even though the determined
offset was typically less than half a pixel.  Cosmic-ray hit removal was
implemented within \texttt{p3d\_cobjex}, utilizing the
\texttt{PyCosmic} \citep{Husemann2012} algorithm. We also chose the option
to subtract a scattered light model before extraction, as
advised for PMAS LArr data \citep{Sandin2012}. Next, the
\texttt{p3d\_sensfunc} routine was utilized to create sensitivity
functions. Here we first created a 1D standard star spectra with the
\texttt{p3d\_sensfunc} GUI using the extracted standard star
observations. Then these sensitivity functions were applied to every
extracted target exposure using the \texttt{p3d\_fluxcal} task.
For correction of atmospheric extinction we used values from the
empirical Calar Alto extinction curve model \citep{Sanchez2007a}.

The final data products resulting from the \texttt{p3d} pipeline are
flux calibrated datacubes for all target exposures.
\texttt{p3d} also produces the corresponding error cubes, containing
the standard-deviation of each volume pixel (`voxel').  After trimming regions affected
by vignetting on the detector \citep{Roth2010}, the cubes cover a
wavelength range from 3600\,--\,4600\,\AA{} sampled on identical
wavelength grids ($\Delta \lambda
=0.75\,\mathrm{\AA{}}\,\mathrm{px}^{-1}$). Their spectral resolution,
determined from fitting 1D Gaussian profiles to several lines in the
spectra of the extracted arc-lamp frames, is $v_\mathrm{FWHM}
\approx160$ km\,s$^{-1}$ ($R\approx1850$).  We note there is a spatial
and spectral dependence of $R$ in PMAS \citep[see
also][Sect. 6.2]{Sanchez2012}, and the value reported here is the median
near the centrum of our wavelength range.

\subsection{Sky subtraction}
\label{sec:sky-subtraction}

To remove night sky emission from the datacubes, we created for each
target a median spectrum from the ring of spaxels bracketing the FoV
(ignoring low-transmission fibres) and subtracted this from every
spaxel.  Even though the median is more robust against contamination
of signal from the target than the mean, it is still possible that we
subtracted a fraction of the nebular emission if this emission
extended out into and beyond the FoV.  However, we note that most of
the known extended Ly$\alpha$ regions\footnote{Using the compilation
  in Fig.~3 of \cite{Cantalupo2014} and 10 additional detections
  reported by \cite{Hennawi2013}.}  around RQQs have projected maximum
extents of $\lesssim 65\,$kpc, corresponding to $\lesssim 8$\arcsec\
(the PMAS FoV) at the redshifts of our objects.  Nevertheless, we
first visually inspected all datacubes if any obvious extended
emission features were present, and then checked the individual
outer-ring median spectra if they contain spikes not attributable to
sky lines. Both tests were negative, so we are certain at this point
not having accidentally removed very bright extended nebular
emission. Nevertheless, we will return to this point in our analysis
in Sect.~\ref{sec:effect-sky-subtr}, showing that for special cases of
large-scale extended emission our observing strategy may have been
less than optimal.

\subsection{Stacking the individual exposures for each QSO}
\label{sec:stacking}

\begin{figure*}
  \centering
  \includegraphics{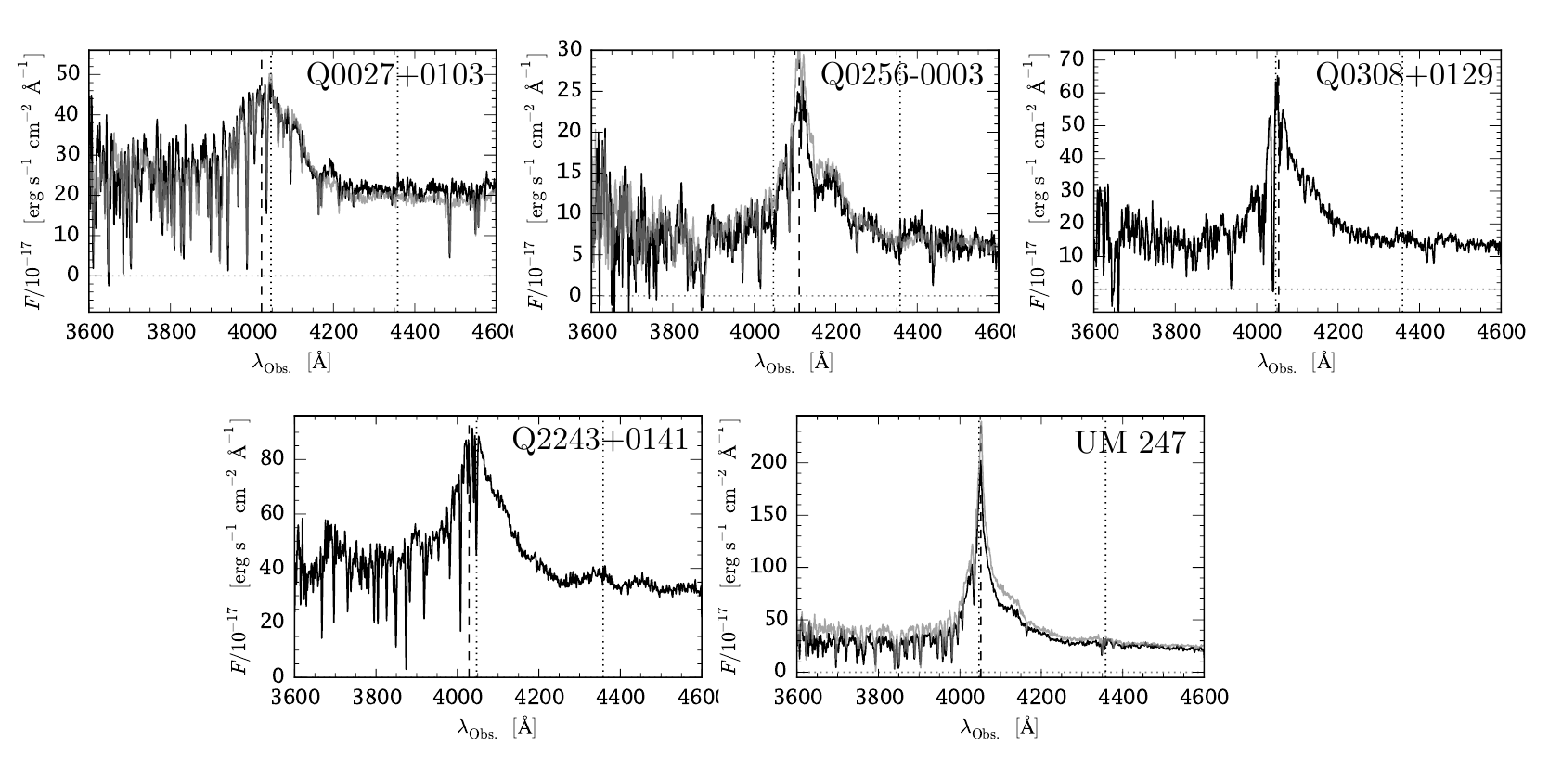}\vspace{-1em}
   \caption{Extracted spectra from the PMAS datacubes in a 3\arcsec
     diameter aperture. For those objects where available
     (Q0027$+$0103, Q0256$-$0003 \& UM 247) SDSS DR9 spectra have
     been  over-plotted  in grey to illustrate the quality of our flux
     calibration. The vertical dashed line shows the wavelength of
     Ly$\alpha$ at the QSOs redshift and the vertical dotted lines
     indicate the wavelengths of the artificial telluric \ion{Hg}{i}
     ($4047\,\mbox{\AA{}}$ \& $4358\,\mbox{\AA{}}$) emission-lines.
   } 
  \label{fig:spectra}
\end{figure*}

To increase the signal-to-noise of faint emission we stacked the
individual sky-subtracted \& flux-calibrated datacubes.  During the
observations the QSO was centered in the LArr FoV using the
acquisition and guiding (A\&G) system of PMAS. No dithering scheme was
intended, to avoid cross-correlation of neighbouring spaxels on the
sky.  Unfortunately the filter wheel of the A\&G camera was set to
V-Band, causing the QSOs centroid point source to move around within
our bluer wavelength range due to differential atmospheric refraction,
even though the guide star observed with the A\&G system remained
steady.  We corrected for these unintended dither offsets by spatial
integer pixel shifts before stacking. Integer pixel shifts were
preferred over fractional shifts to avoid interpolation effects and
thus keeping the original observational information as unaffected as
possible.

In practice, we first determined the QSOs centroid position in the
individual exposures using images created from summing up datacube
layers around $\lambda_\mathrm{Ly\alpha}(z_\mathrm{QSO})$. If a
centroid was shifted more than a half pixel along a spatial axis, this
cube was then shifted by a pixel along this axis in the opposite
direction. No exposure needed to be shifted by more than one pixel in
$x$ or $y$ direction.  Because of the varying sky background (and
different exposure times in the case of Q0256$-$0003) we employed a
variance-weighted mean for stacking, using the error cubes generated
by the pipeline.

Having reduced the raw data as described, we were now left with five
datacubes containing voxels $F_{x,y,z}$, where $(x,y)$ are the spatial
indices, and $z$ is the index for the spectral layers, in units of
$10^{-17}\,\mathrm{erg}\,\mathrm{s}^{-1}\mathrm{cm}^{-2}\mathrm{\AA{}}$.

\section{Analysis \& results}
\label{sec:analysis}

\subsection{QSO spectra}
\label{sec:1d-qso-spectra}

We show spectra of our five QSOs in Fig.~\ref{fig:spectra}. These spectra
were extracted in a 3\arcsec{} diameter aperture centered on the
highest S/N spectral pixel (spaxel).  Where
available, we overlay SDSS DR9 \citep{Ahn2012} spectra. Our spectra
differ to SDSS by $\lesssim$ 10 \% for Q0027$+$0103 \& Q0308$+$0129;
for UM\,247 there is a systematically offset by $\lesssim$ 20 \% in the
blue. Quasar variability, spectrophotometric uncertainties in
the SDSS or imperfections in our flux calibration could be
reasons for these differences.  We note that for Q0308$+$0129 the
observing conditions were not photometric, thus our fluxes of this object
are likely somewhat too low.

The night sky emission line spectrum in Calar Alto shows significant
man-made contributions, arising from tropospheric scattering of
high-pressure street lamps in nearby populated areas.  In our
wavelength range the \ion{Hg}{i} $4047\,\mbox{\AA{}}$ \&
$4358\,\mbox{\AA{}}$ emission lines are prominent.  Wavelengths of
these lines and the expected Ly$\alpha$ wavelength
$\lambda^{\mathrm{Obs.}}_\mathrm{Ly\alpha} = (1+z_\mathrm{QSO}) \cdot
\lambda_\mathrm{Ly\alpha}$
are indicated in Fig.~\ref{fig:spectra}. As can be seen, the
$\lambda4047$ line is located unfavorably close near the expected peak
of the Ly$\alpha$ emission for 4 of our QSOs. We had not
  anticipated the strength of this anthropic line in the planning of
  our observations. This unresolved line has an average surface
brightness of
$F^\text{\ion{Hg}{i}}_{4047}=3.5\times10^{-16}\,\mathrm{erg}\,\mathrm{s}^{-1}\mathrm{cm}^{-2}\mathrm{arcsec}^{-2}$
\citep{Sanchez2007a} and thus, by amplifying the background noise in a
narrow-band window around $\lambda^{\mathrm{Obs.}}_\mathrm{Ly\alpha}$,
contributes negatively to our efforts in uncovering extended
Ly$\alpha$ emission.  The sensitivity in this narrow-band is further
decreased by residuals from sky subtraction, resulting from the
varying spectral resolution across the FoV
(Sect.~\ref{sec:datar-with-p3d}).

\subsection{Subtraction of QSO emission from the datacubes}
\label{sec:psf}

Due to the PSF broadening, spectra at radial distances $\lesssim
2$\arcsec{} from the QSO center are visibly contaminated by the
nuclear spectrum.  Thus, in order to reveal possible extended
Ly$\alpha$ emission around the QSO, we had to remove this
contamination. \citetalias{Christensen2006} experimented with
three algorithms to achieve this \emph{deblending} in PMAS LArr
datacubes:
\begin{itemize}
\item \emph{On- minus scaled off-band method}: An on- and off-band
  image is created from the datacube. The latter is then scaled and
  subtracted from the former, thereby revealing possible contribution
  of extended emission in the on-band.  Here the \emph{on-band} is obtained as a
  summation in wavelength direction over all datacube layers expected to
  contain the extended-emission signal, while the \emph{off-band} is a
  sum over nearby layers not being affected by extended
  emission. Asserting that the extended emission contributes only a
  few percent to the on-bands peak intensity (i.e. QSO$+$extended
  emission), the ratio between the peak intensities in on- and off-band is
  a sufficient scale factor for the subtraction.
\item \emph{Analytical PSF Extraction}: An analytical PSF model is
  fitted and subtracted from every datacube layer, thereby creating a
  datacube which should only contain non-nuclear emission. Then
  an iterative scheme is employed, where in a second iteration
  positional and shape parameters of the PSF function are fixed, using
  values from the first iteration with the constraint of allowing them
  to vary only smoothly with wavelength \citep[see e.g.][for more
  elaborate versions of this technique involving multiple point-sources
  within the FOV]{Wisotzki2003,Kamann2013}. According to
  \citetalias{Christensen2006} the analytical PSF extraction produces
  unsatisfying results in PMAS in terms of QSO residuals, so we did not consider
  applying this method to our data.
\item \emph{Empirical PSF Subtraction}: By summation over datacube
  layers expected to be unaffected by extended emission, an empirical
  PSF image is created. It is then scaled and subtracted from every
  layer, thus creating a residual datacube containing extended
  emission. To find the scaling the PSF image with respect to the datacube layers, 
  a spectrum is aperture-extracted from the datacube, with every spaxel
  in the aperture having weights assigned via the PSF image.  An
  iterative scheme is employed to remove contamination in the scaling
  spectrum from extended emission possibly covering parts of the
  aperture: the datacube resulting from the deblending is subtracted
  from the undeblended datacube from which in turn the next scaling
  spectrum is extracted.  A variant of this method was developed by
  \cite{Husemann2013a}, where in subsequent iterations the correction
  is achieved using an annulus-extracted spectrum from the deblended
  datacube.
\end{itemize}

We emphasize the methodological similarity between the
\emph{on- minus scaled off-band method} and the \emph{empirical PSF
  subtraction} deblending: If the PSF image in the latter is created
from the same layers as the off-band in the former, every layer in the
deblended datacube can be thought of as an `on-band' having a scaled
`off-band' subtracted. If, moreover, no iterative corrections are
applied, the image resulting from the \emph{on- minus scaled off-band
  method} is identical to the image resulting from summation over the
on-band defining layers in the empirical PSF subtracted datacube. Note
that the iterative correction only increases the fidelity of an
extended emission signal when it is measurably present after the first
deblending iteration; otherwise only noise is shuffled around in subsequent
iterations.

Because of this similarity we applied solely the
\emph{empirical PSF subtraction} method to our datacubes.  As PSF
image we take $\sim40$ layers ($\sim30$~\AA{}) beginning $\sim$10~\AA{}
redwards of $\lambda_{\mathrm{Ly\alpha}}(z_\mathrm{QSO})$. We found
this choice to produce the best results in terms of subtraction
residuals, because of the following reasons: Firstly, for the
red side of $\lambda_{\mathrm{Ly\alpha}}(z_\mathrm{QSO})$, a redward
band gives a better signal-to-noise ratio for the PSF image due to
$z_\mathrm{abs} \approx z_\mathrm{em}$ Ly$\alpha$ absorbers appearing
blueward of $\lambda_{\mathrm{Ly\alpha}}(z_\mathrm{QSO})$ (see
Fig.~\ref{fig:spectra}). Secondly, as we did not correct for
differential atmospheric refraction, the position and shape of the PSF
changes with $\lambda$, thus selecting layers further away would
produce a non-optimal representation of the PSF at the expected
position of Ly$\alpha$ fuzz, resulting in stronger residuals from the
subtraction ($\lambda_{\mathrm{Ly\alpha}}(z_\mathrm{QSO})$). This is
also the reason why we chose the spectral window to be relatively narrow. 

We experimented with different extraction apertures for the scaling spectrum.  
Visual inspection revealed no extended emission after the first iteration in all
objects.  As expected (see above), performing more iterations does not
improve upon this.  We thus decided to use the smallest possible
aperture for scaling, i.e. the single spaxel with the highest S/N. 
Despite this, we could not find any evidence for extended Ly$\alpha$ fuzz
around any of our targets, except a small-scale feature to the north of UM\,247.

To visualize our findings, we show in Fig.~\ref{fig:residual_images}
narrow-band images centered on
$\lambda_{\mathrm{Ly}\alpha}(z_\mathrm{QSO})$ created by summing over
20 layers (15\AA{}) of the residual datacubes. Instead of using
physical units, the scale on the colour bar uses multiples of the
standard-deviation per pixel. This colour bar scaling simplifies
the judgment whether features seen in this image are actually
significant, and it also makes the comparison between the panels straightforward. 
Except for UM\,247, none of these images shows any significant feature.

\begin{figure*}
  \centering
      \begin{tikzpicture}
      \node at (0,0)
      {\includegraphics[scale=0.45]{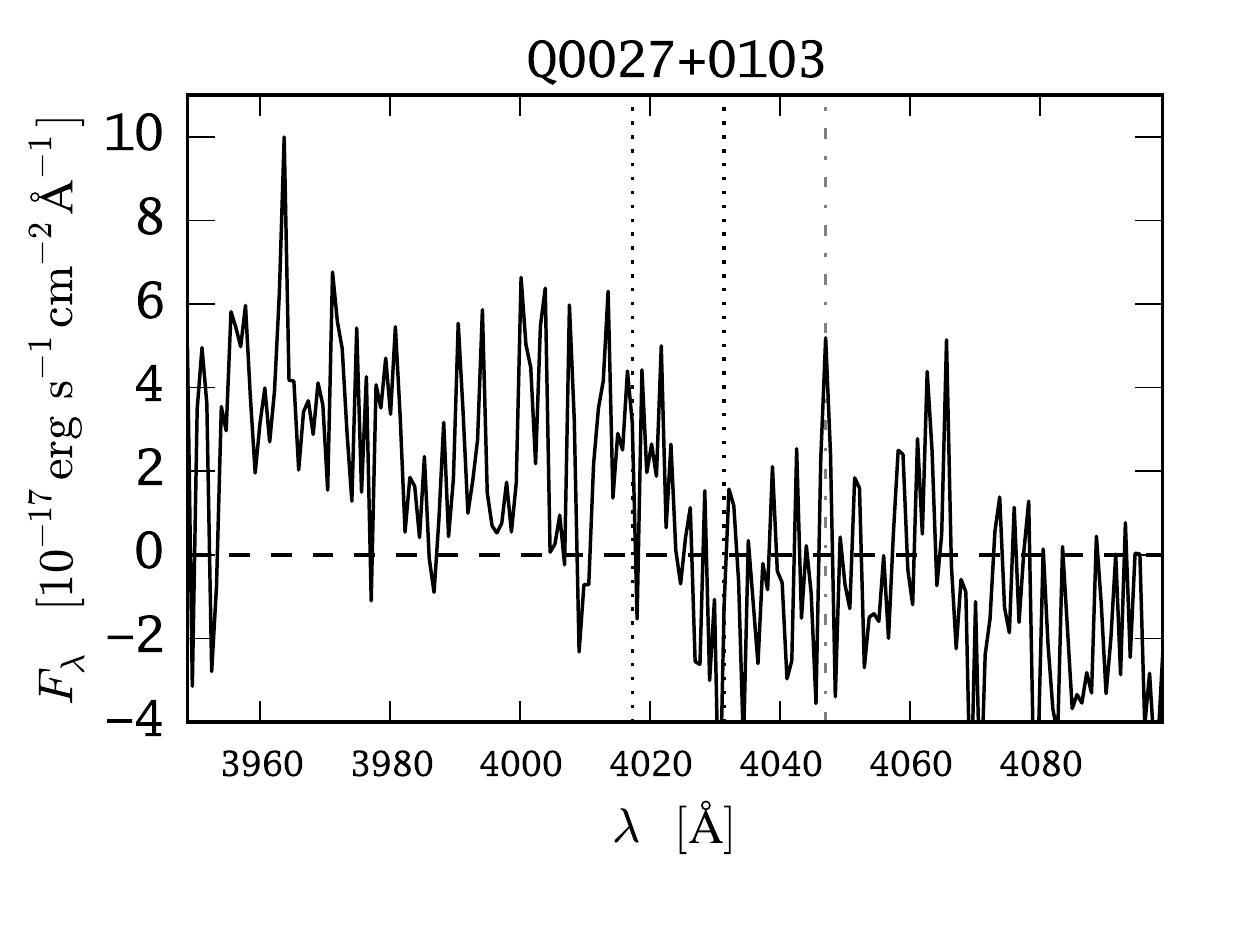}};
      \node at (5.5,0)
      {\includegraphics[scale=0.45]{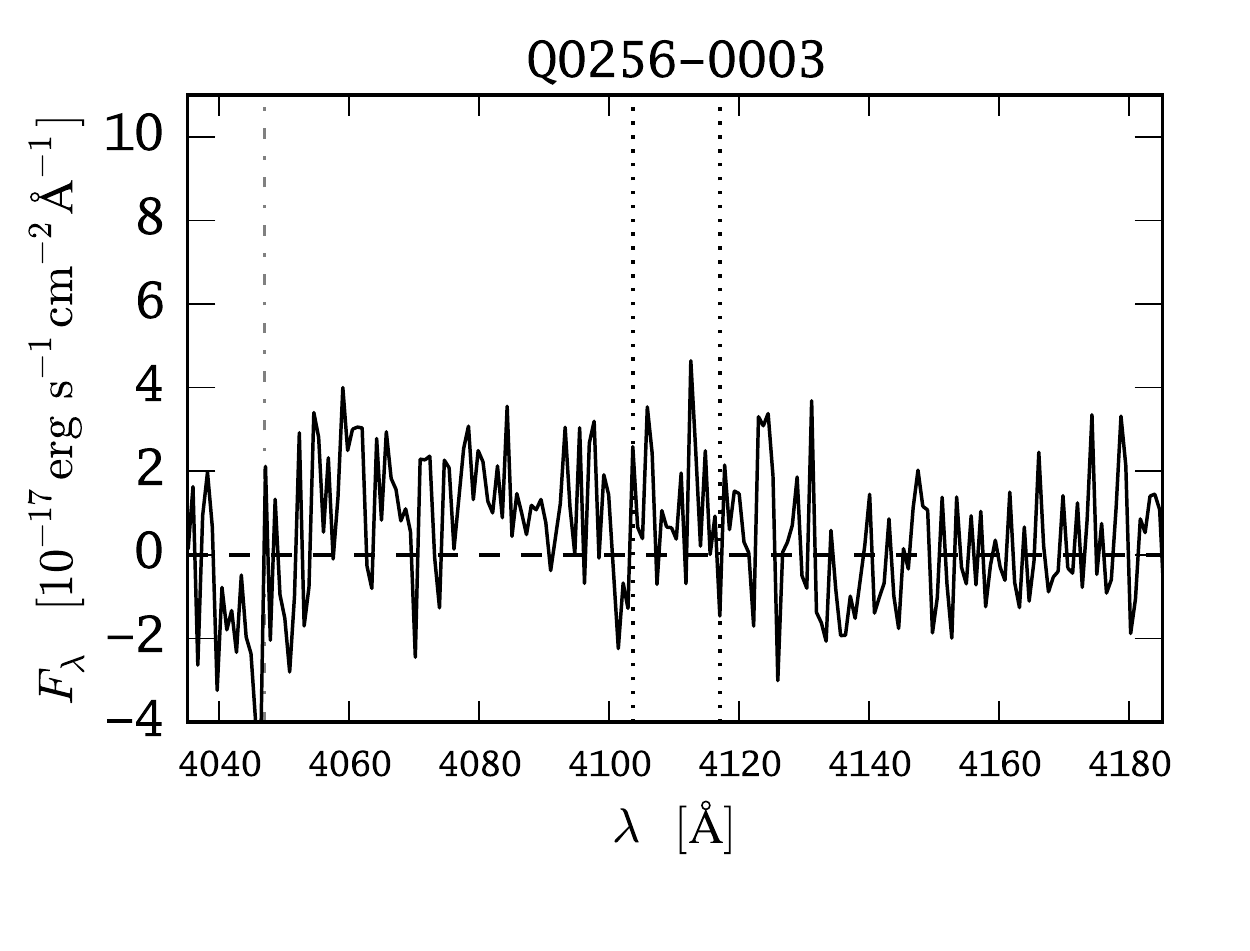}};
      \node at (11,0)
      {\includegraphics[scale=0.45]{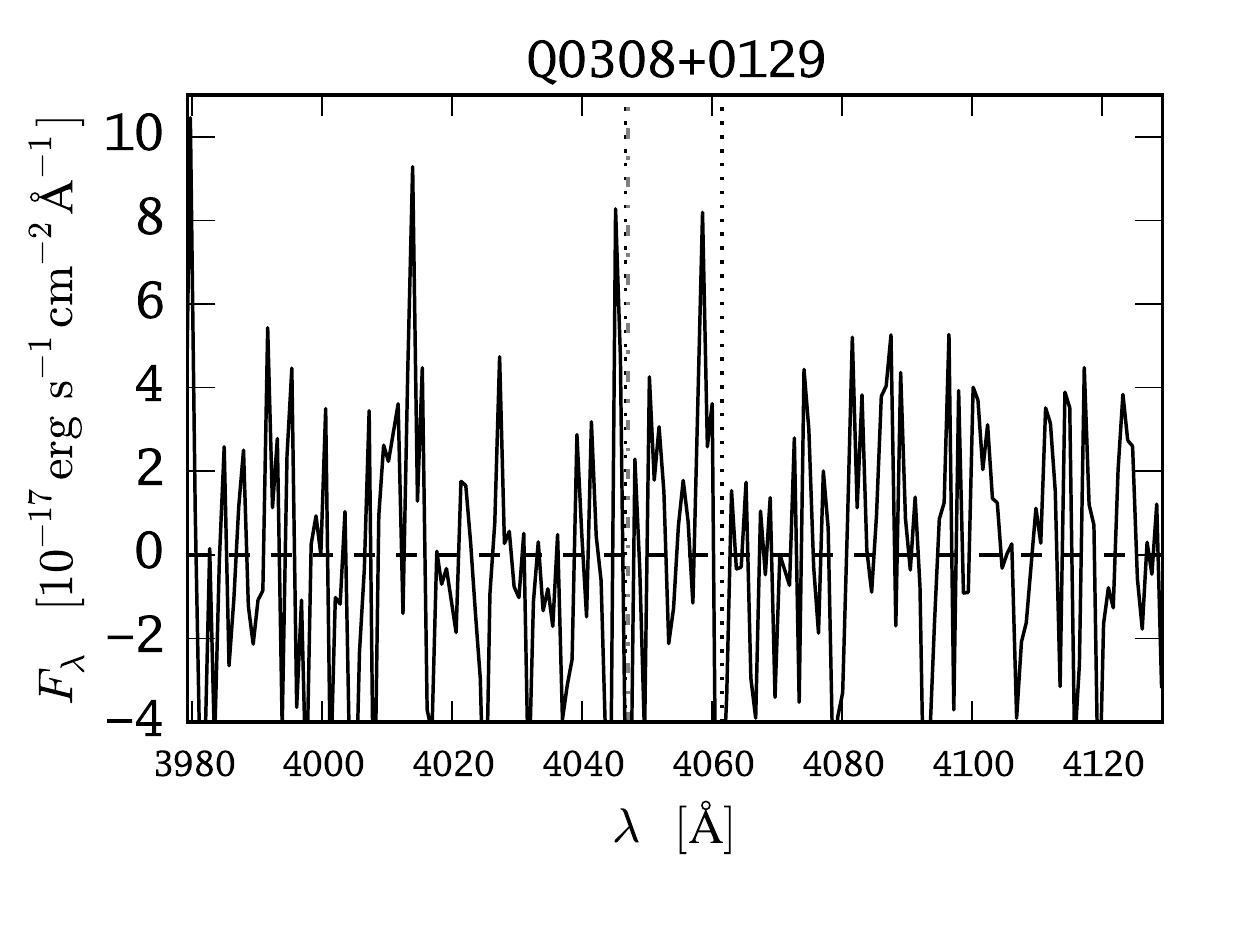}};
      \node at (2.25,-4)
      {\includegraphics[scale=0.45]{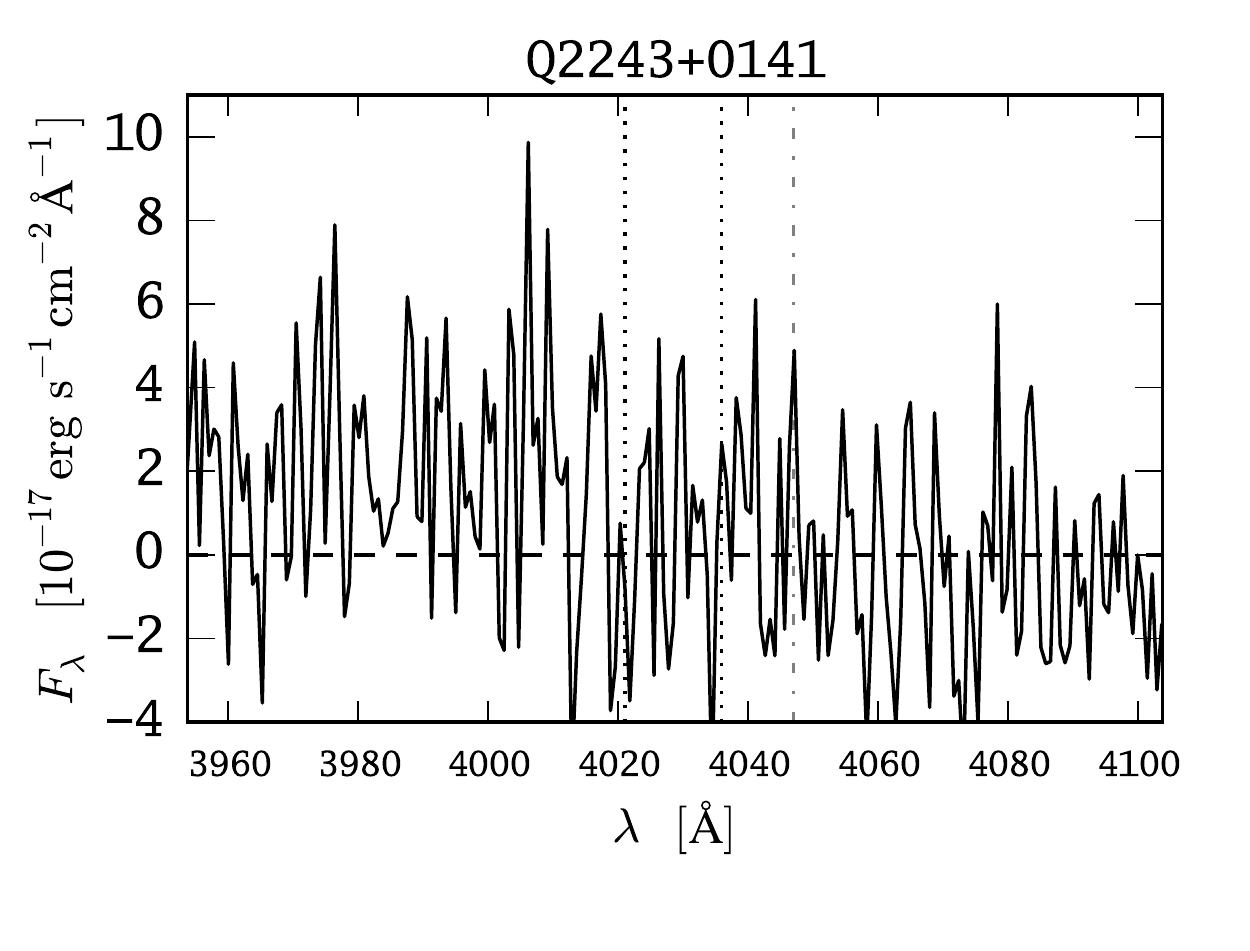}};
      \node at (7.75,-4)
      {{\includegraphics[scale=0.45]{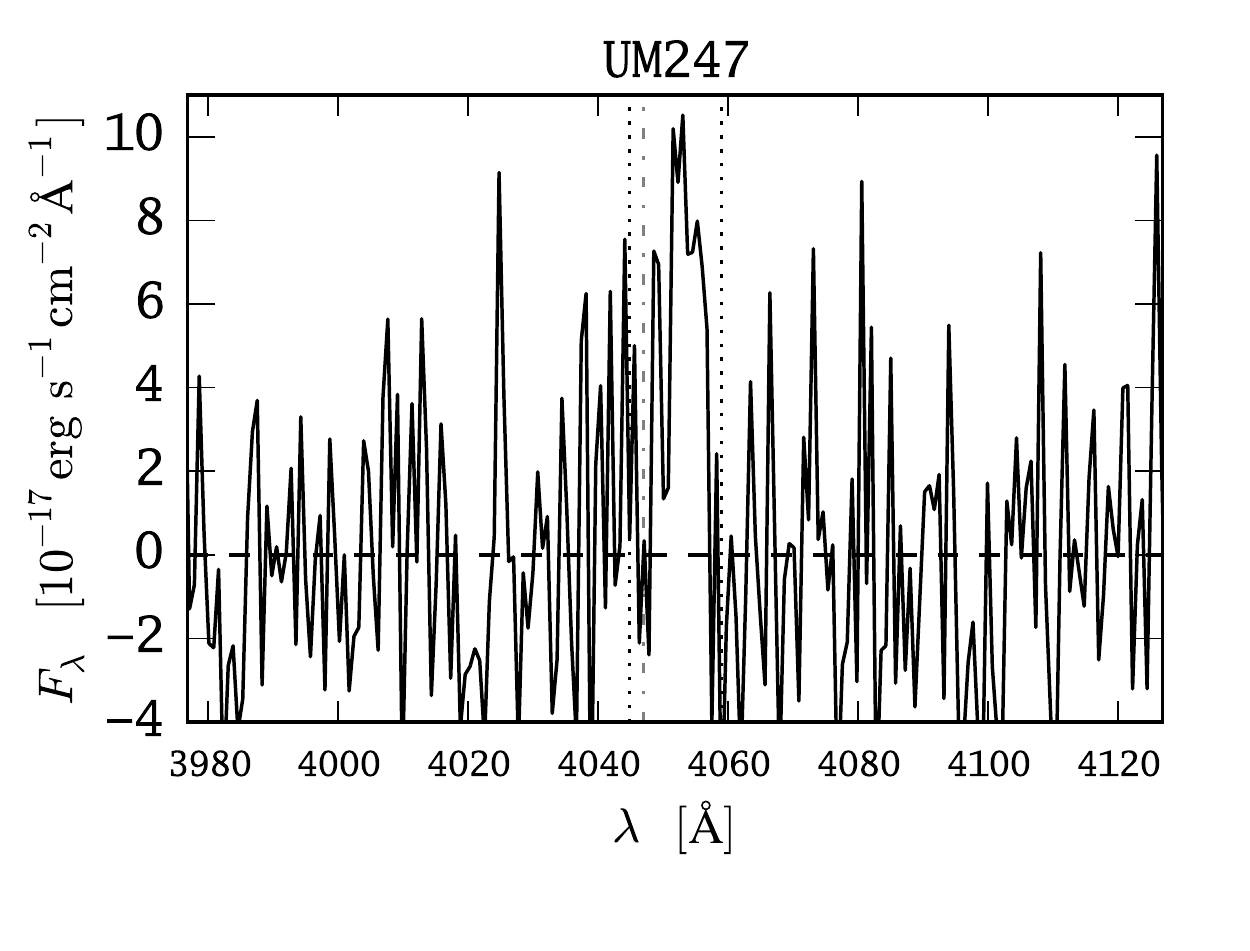}}};
    \end{tikzpicture}
    \vspace{-1em}
    \caption{Residual spectra for our 5 QSOs from the QSO-subtracted
      datacubes, extracted within the $r_c=1.25$\arcsec{} aperture
      (cf. Sect.~\ref{sec:calc-detect-limits}) around the scaling
      spectrum. The \emph{vertical dotted lines} indicate the
      boundaries of the narrow-band image (15 \AA{} $=$ 20 layers),
      shown in Fig.~\ref{fig:residual_images}. The \emph{vertical
      dashed-dotted lines} show the position of the \ion{Hg}{I}-sky
      line (Sect.~\ref{sec:1d-qso-spectra}).}
  \label{fig:residual_spec_plots}
\end{figure*}

\begin{figure*}
  \centering
  \includegraphics[width=0.85\textwidth]{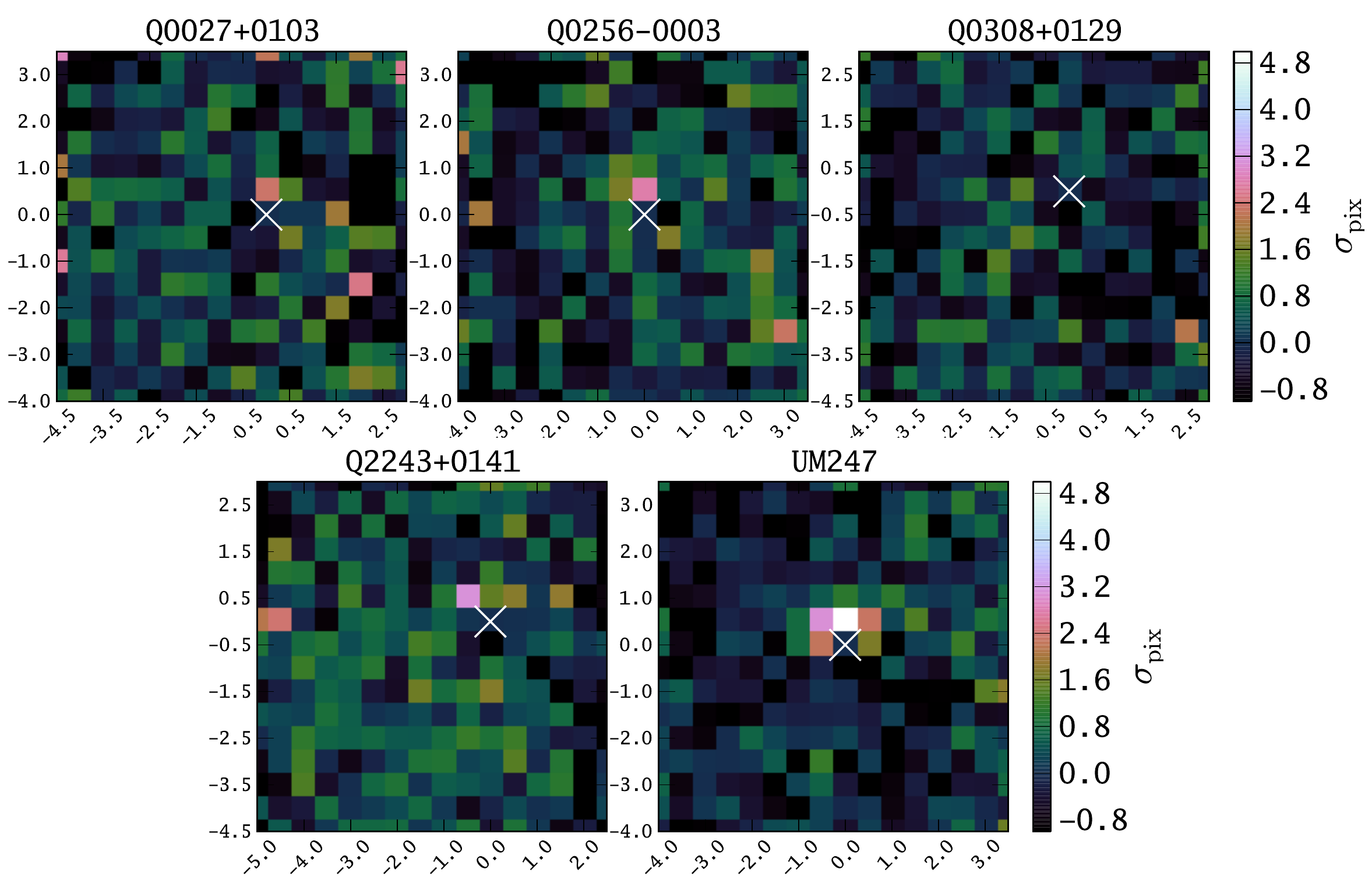}
  \caption{Narrow-band images (15\AA{} wide, centered on
    $\lambda_{\mathrm{Ly}\alpha}(z_\mathrm{QSO})$) for our 5 QSOs.
    These images were created from the QSO-subtracted datacubes using
    the empirical PSF subtraction method. The position of the spaxel
    used for scaling the PSF image is indicated by a cross;
    this position corresponds to the centroid of the
    QSO nucleus. Colours indicate multiples
    of the standard-deviation per pixel $\sigma_\mathrm{pix}$,
    calculated for each image. The values of $\sigma_\mathrm{pix}$
    are (2.3, 1.4, 2.7, 2.6, 3.5)$\times 10^{-17}$erg$\,$s$^{-1}$\,cm$^{-2}$,
    from top left to bottom right.  North is up, and East is to the
    left. Axes ticks are $\Delta \delta$ and $\Delta \alpha$ in
    arcseconds with respect to the QSO centroid. }
  \label{fig:residual_images}
\end{figure*}

In Fig.~\ref{fig:residual_spec_plots} we show spectra from the
residual datacubes extracted within a small circular aperture
consisting of 20 spaxels (aperture with outer radius
$r_C=1.25$\arcsec{}, cf. Sect.~\ref{sec:calc-detect-limits}) centered around
the scaling spectrum. Again, only UM\,247 displays a distinct spectral line
that seems inconsistent with noise. We discuss this feature further 
in Sect.~\ref{sec:small-scale-lyalpha}.

\subsection{Estimation of detection limits}
\label{sec:calc-detect-limits}

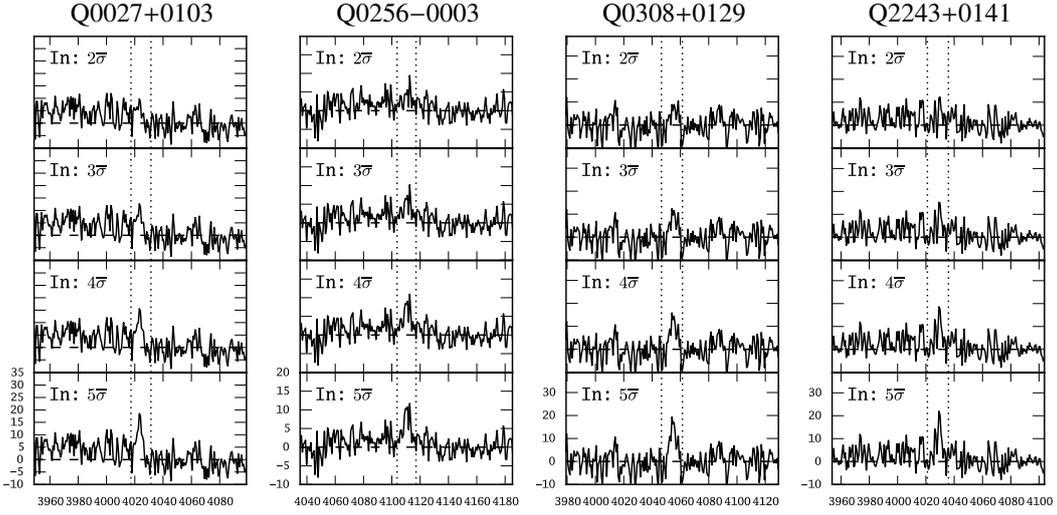
\begin{figure*}
\centering
\begin{tikzpicture}
  \node at (0.2,3.4) {\color{black} Q0027$+$0103};
  \node at (0,0)
  {\includegraphics[scale=0.55]{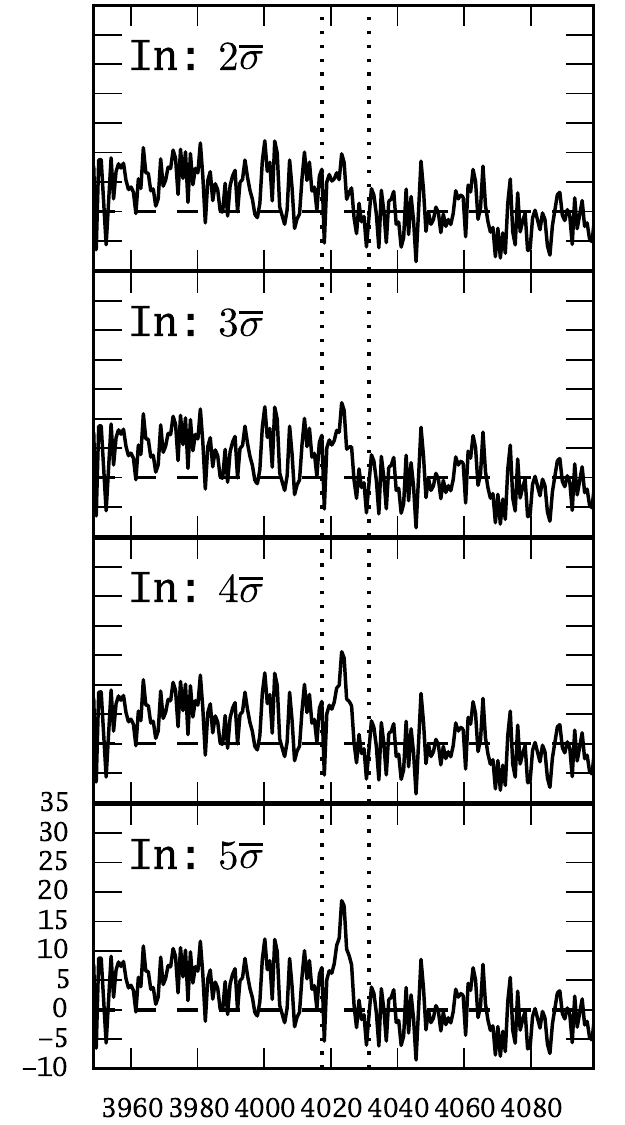}};
  \node at (3.7,3.4) {\color{black} Q0256$-$0003};
  \node at (3.5,0)
  {\includegraphics[scale=0.55]{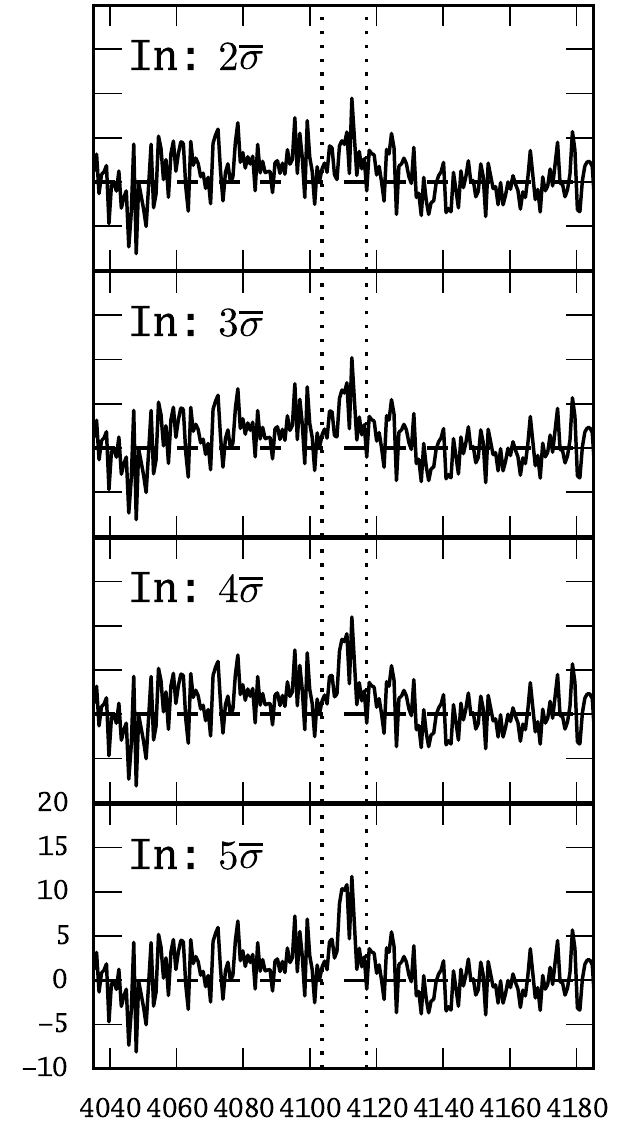}};
  \node at (7.2,3.4) {\color{black} Q0308$+$0129};
  \node at (7,0)
  {\includegraphics[scale=0.55]{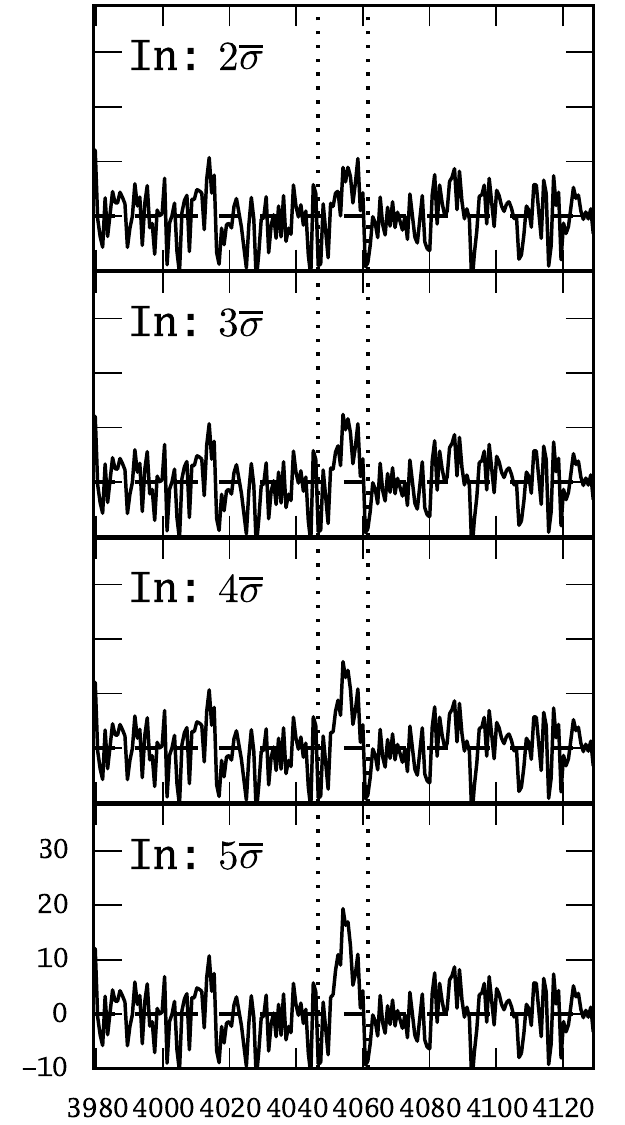}}; 
  \node at (10.7,3.4) {\color{black} Q2243$+$0141};
  \node at (10.5,0)
  {\includegraphics[scale=0.55]{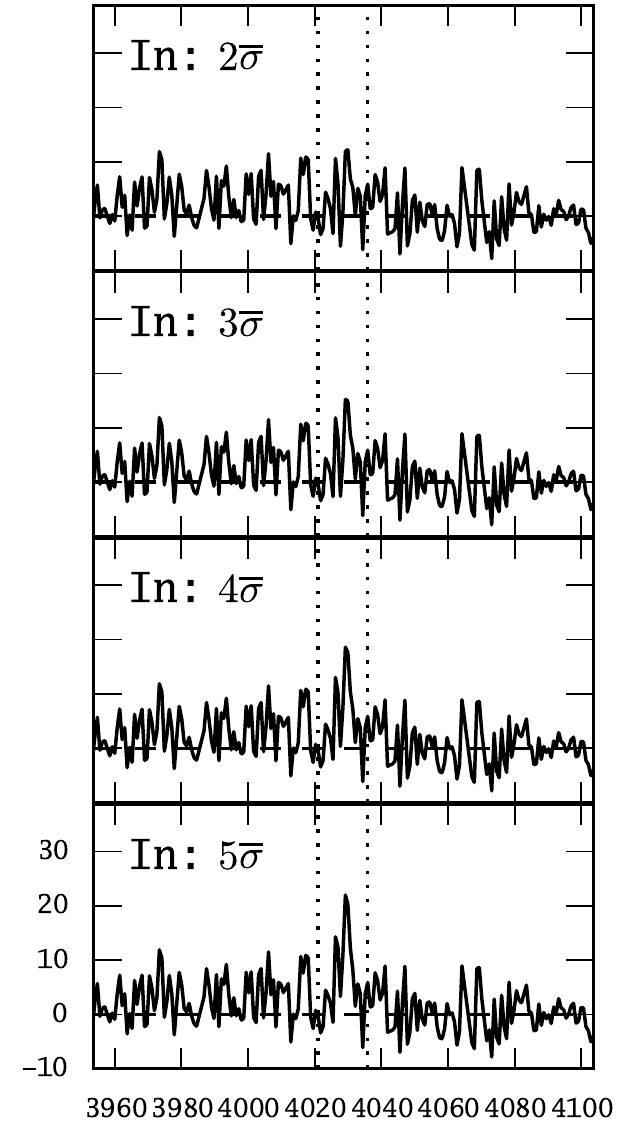}};
\end{tikzpicture}
\caption{Examples of recovered spectra after adding artificial nebulae
  into the datacube prior to the empirical PSF subtraction.  The
  surface brightness of the artificial nebulae was scaled with integer
  values of $n_\mathrm{spatial}$ according to Eq.~(\ref{eq:4}). After
  empirical PSF subtraction the spectrum was extracted from the
  residual datacube using the $r_C=1.25$\arcsec{} aperture (i.e. the
  same as in Fig.~\ref{fig:residual_spec_plots}). The spectra are
  shown in units of $10^{-17}$erg\,s$^{-1}$cm$^{-2}$\AA{}$^{-1}$. Note
  that artificial nebulae with $n_\mathrm{spat}\ge 5$ can be 
  unambiguously discriminated from the background for all objects.}
  \label{fig:detection_sim}
\end{figure*}

\begin{figure}
  \begin{center}
    \includegraphics[width=0.45\textwidth]{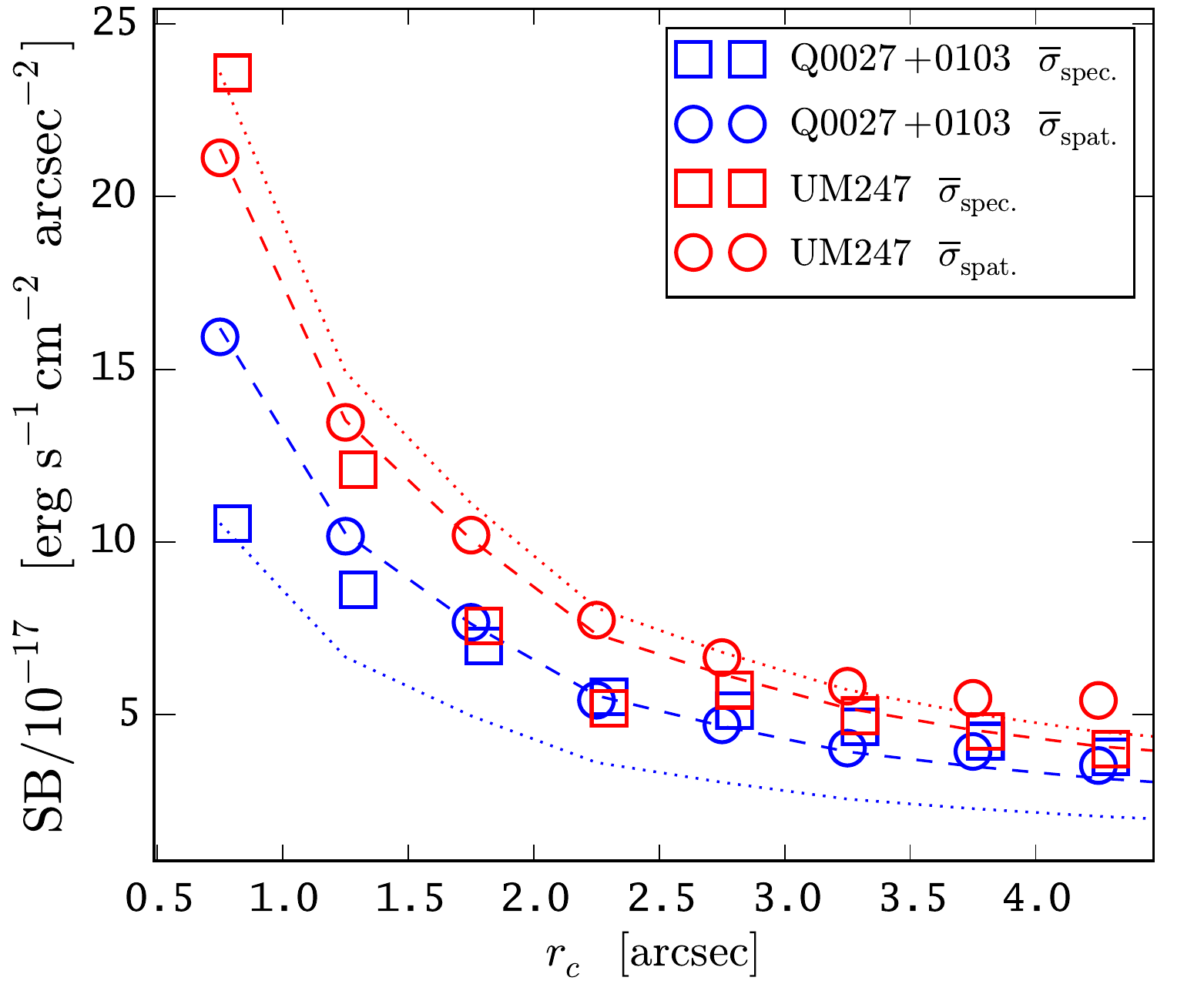}
  \end{center}
  \vspace{-1em} 
  \caption{Comparison of formal 5$\sigma$ surface brightness detection
    limits as a function of aperture radius using the different noise
    estimators $\overline{\sigma}_\mathrm{spat}$ and
    $\overline{\sigma}_\mathrm{spec}$, exemplarily shown for two objects
    (Q0027$+$0103 - \emph{blue} symbols, UM\,247 - \emph{red} symbols
    panel). The agreement between $\overline{\sigma}_\mathrm{spat}$
    and $\overline{\sigma}_\mathrm{spec}$ is similarly good for the other
    objects.}
  \label{fig:detsigexample}
\end{figure}

\begin{figure*}
  \centering\vspace{-0.5em}
  \includegraphics[width=0.33\textwidth]{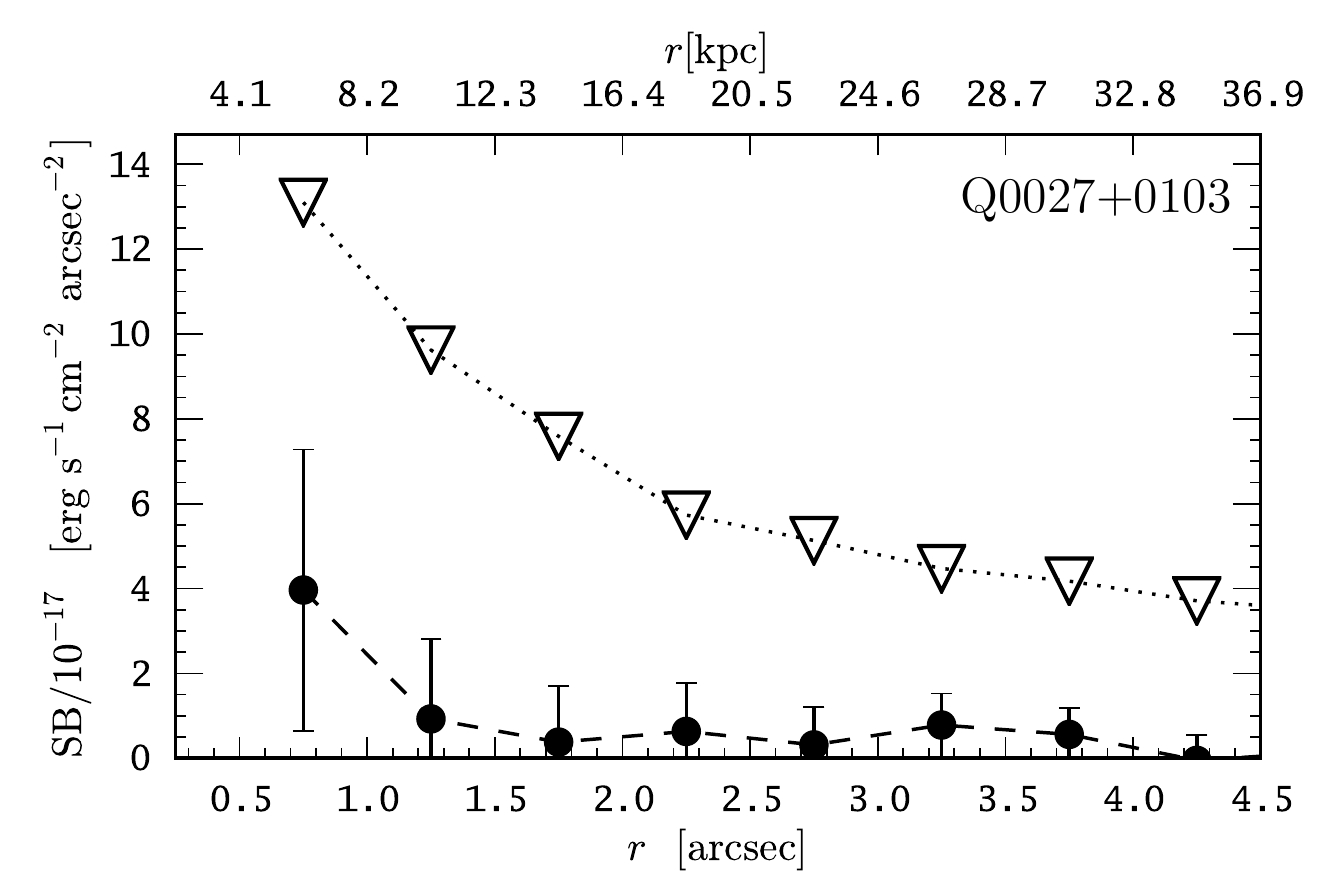}
  \includegraphics[width=0.33\textwidth]{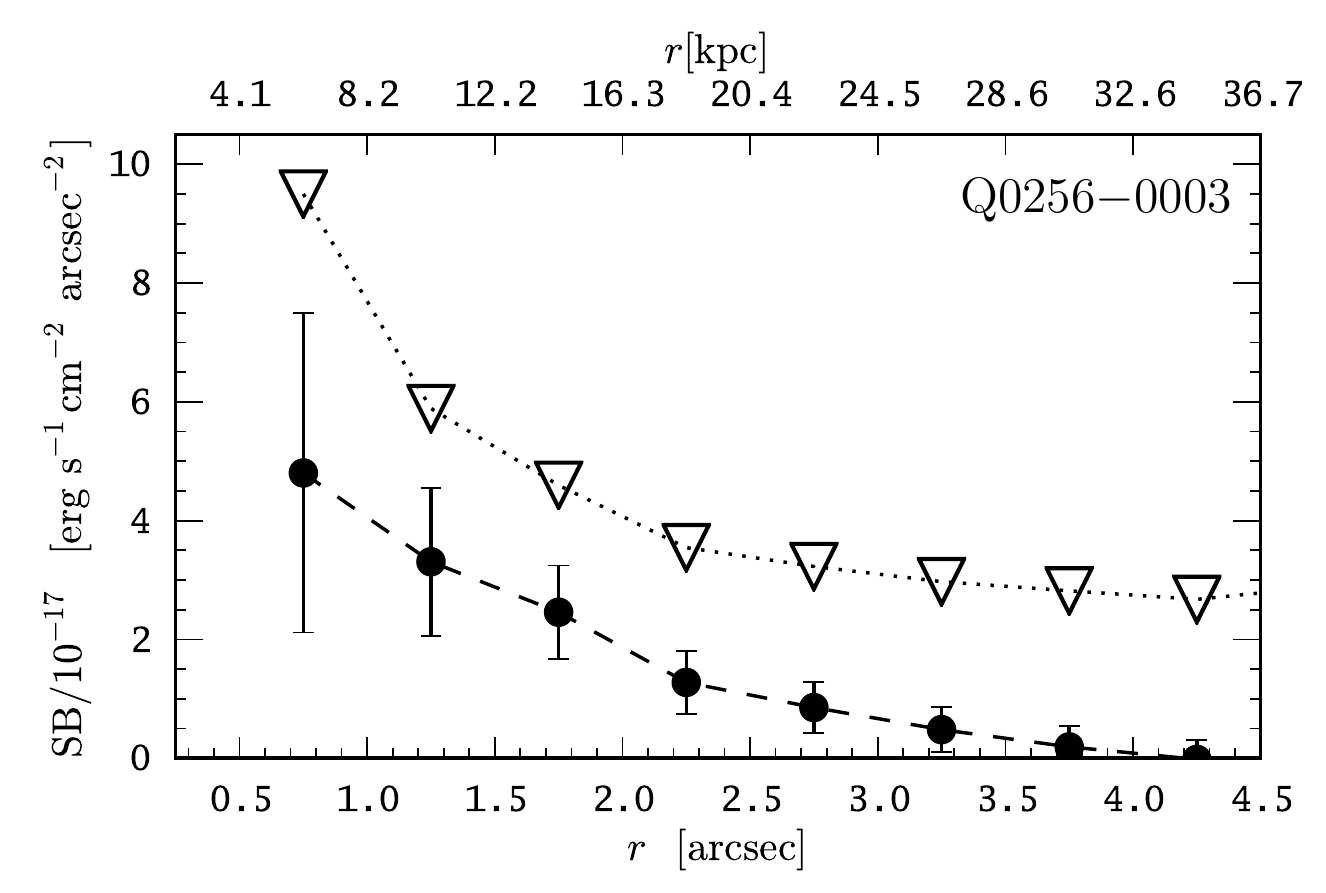}
  \includegraphics[width=0.33\textwidth]{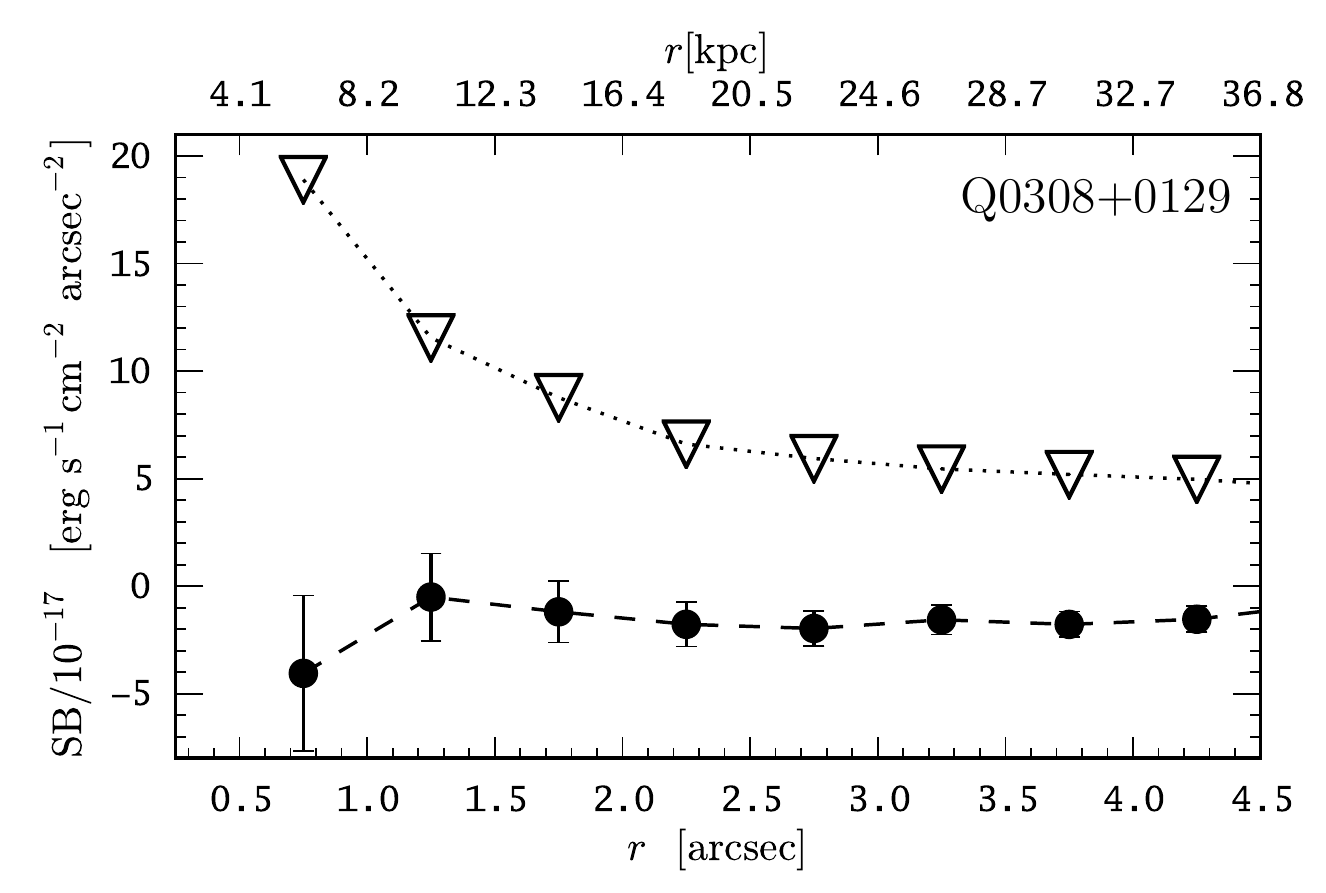}\vspace{-0.5em}
  \includegraphics[width=0.33\textwidth]{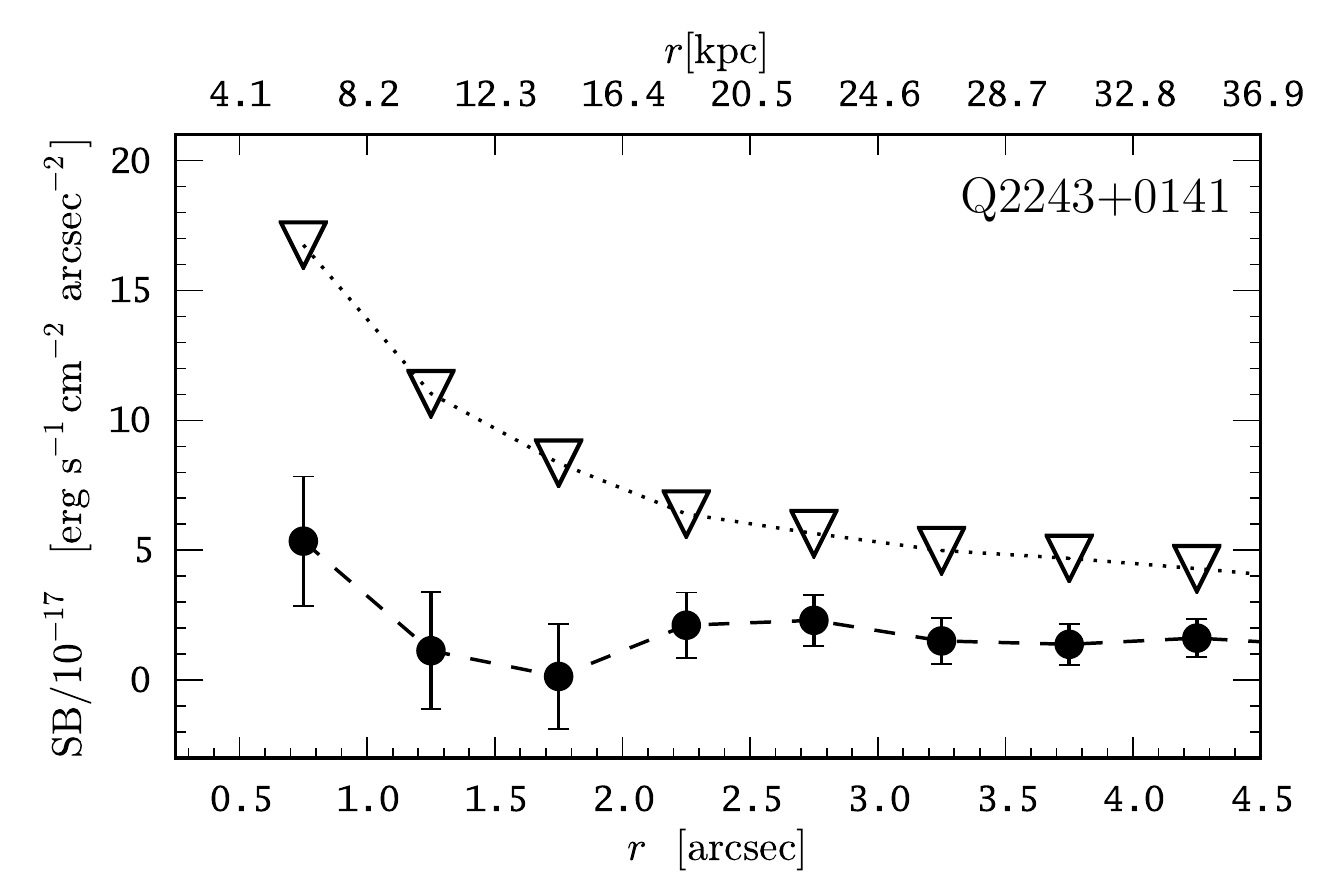}
  \includegraphics[width=0.33\textwidth]{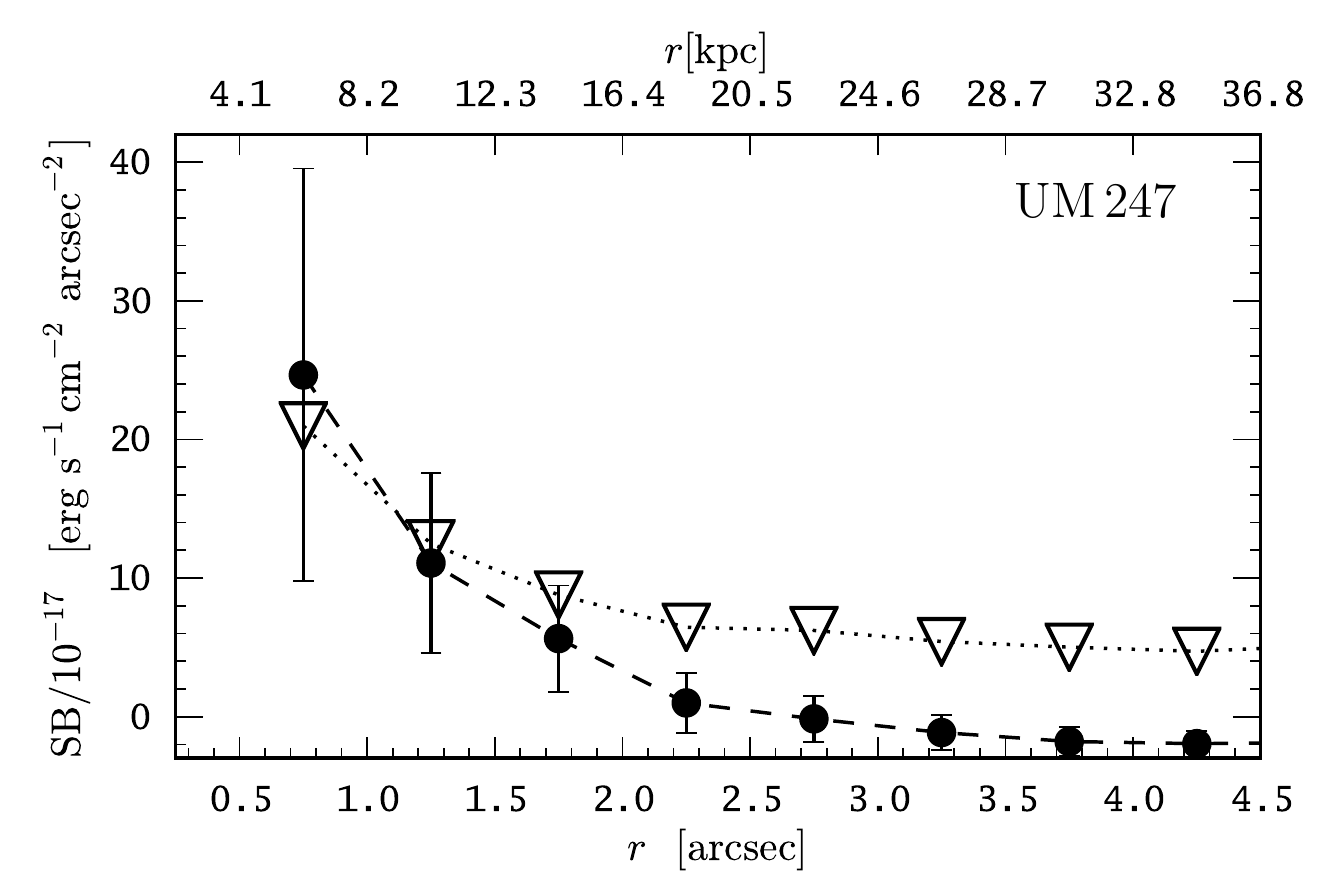}
  \caption{Surface brightness upper limits (triangles) for circularly symmetric
    extended Ly$\alpha$ emission around our 5 QSOs as a function of
    radial angular (bottom abscissa) or physical extent (top
    abscissa). The integrated observed signal within each
    apertures is shown by the black points, where the error
    bars indicates the standard deviation within the aperture, as a
    measure of the irregularity of the flux distribution within the
    aperture.}
  \label{fig:radprofonepanel}
\end{figure*}

From our visual inspection of the residual datacubes (see also
Figs.~\ref{fig:residual_images}~\&~\ref{fig:residual_spec_plots}) we
conclude that in 4 out of 5 objects we have no significant
detections. We now want to constrain an upper limit in surface
brightness for extended Ly$\alpha$ nebulae around those QSOs. To do
so, we define confidence limits for rejecting the null hypothesis, 
which states that there is no extended Ly$\alpha$ emission present in the data. 
The upper limit is then given as the minimum surface brightness for which we would not
be able to confidently reject the null hypothesis anymore.

We adopted circular apertures centered on the scaling spectrum. 
In absence of morphological information for the
non-detections, this simplification appears reasonable, although it is
known that QSO CGM Ly$\alpha$ nebulae can be asymmetric
\citep[e.g.][]{Weidinger2004,Rauch2013a}. For such nebulae the average
surface brightness within a circular aperture will generally be lower
than a surface brightness obtained within an isophote. 

We explored successively larger apertures around the central scaling
spectrum by adding annuli with a width of one pixel. This defined 8
``circular'' apertures $C_k$ ($k=1\dots8$) with outer radii of
$r_c=(2k+1)/4$\arcsec{}. The numbers of spaxels $N_C$ in those
apertures are then $N_{C_k} = ($8, 20, 36, 68, 96, 136, 176).

The assessment was performed on the residual datacube $F_{x,y,z}'$ 
after PSF subtraction. The residual signal $S_k$ within the relevant 
spectral layers and inside each circular aperture $C_k$ can be written
as
\begin{equation}
  S_k = \sum_{(x,y)\in
    C_k} I_{x,y} 
  \;\text{,}
\label{eq:1}
\end{equation}
where $I_{x,y} = \sum_{z\in\mathrm{NB}} F_{x,y,z}'$ is the value of a
pixel in the narrow-band images shown in
Fig.~\ref{fig:residual_images}. For the sake of brevity, we do
not explicitly state units and conversions in our equations. 

Assuming for the moment that we had reliable variance
estimates $\sigma_{x,y}^2$ for every pixel of the pseudo narrow-band
image $I_{x,y}$ (or equivalently $\sigma_{x,y,z}^2$ for every voxel of
the residual datacube), the noise in $C_k$ could be written as
\begin{equation}
  \label{eq:2}
  \sigma_k = \frac{1}{N_{C_K}} \sqrt{\sum_{(x,y)\in C_k} \sigma^2_{x,y}} 
\;\text{,}
\end{equation}
Combining now Eqs.~(\ref{eq:1}) and (\ref{eq:2}), we could express the
signal $S_k$ within $C_k$ in amounts of the noise
$\sigma_k$ being present in $C_K$, i.e.
\begin{equation}
\label{eq:3}
S_K = n \cdot
\sigma_k\;\text{,}
\end{equation}
where, under assumption of pure Gaussian noise, $n$ would
directly translate into a probability of the null hypothesis -- no
nebular emission detected -- being false \citep[e.g.][]{Wall1979}. 

However, neither of the above made assumptions -- having reliable variance estimates 
for every voxel and pure Gaussian noise -- are met by our data: The QSO subtraction 
introduces non-Gaussian residuals, as does the background subtraction
(Sect.~\ref{sec:sky-subtraction}), which are not captured by the formal variances. 
There may also be further unknown systematics.
We therefore replaced the formal variances by two empirical proxies for
$\sigma_{x,y}$:
\begin{itemize}
\item $\overline{\sigma}^\mathrm{spat}_k$: The standard deviation
  (relative to an expectation value of zero) per pixel of the narrow-band image $I_{x,y}$
  outside of the circular aperture $C_k$.
\item $\overline{\sigma}^\mathrm{spec}_k$: The standard deviation
  (relative to an expectation value of zero) per spaxel within the circular aperture $C_k$,
  calculated from spectral layers not contributing to the narrow-band images.
  The calculation is limited to wavelengths not further than 75\AA{} away
  from $\lambda(z_\mathrm{QSO})$, since here the empirical PSF
  subtraction starts to produce strong residuals because of
  differential atmospheric refraction. 
\end{itemize}
Note that both noise estimators depend on the the aperture $C_k$.

With these proxies we then set $\sigma_{x,y} \approx
\overline{\sigma}^\mathrm{spat}_k$ or $\sigma_{x,y} \approx
\overline{\sigma}^\mathrm{spec}_k$ for all $x,y$ in Eq.~(\ref{eq:2})
and obtain
\begin{eqnarray}
  S_k & = & n_\mathrm{spat}  \cdot 
  (\overline{\sigma}^\mathrm{spat}_k / \sqrt{N_{C_k}}) \;\text{,} 
\label{eq:4}
  \\
  S_k & = & n_\mathrm{spec} \cdot (\overline{\sigma}^\mathrm{spec}_k
  /\sqrt{N_{C_k}}) \;\text{.} 
\label{eq:5}
\end{eqnarray}
We show below that $n_\mathrm{spec} \approx
n_\mathrm{spat}$ (or equivalently $\overline{\sigma}^\mathrm{spec}_k
\approx \overline{\sigma}^\mathrm{spat}_k$) holds.

The question is, which $n_\mathrm{spec}$ or $n_\mathrm{spat}$ is
required, in each QSO, for a detection?  We addressed this problem by
adding simulated extended emission into our datacubes before
subtracting the QSO. Specifically we used circular nebulae of an
extent that would fill a particular aperture $C_k$.  These simulated
nebulae had a flat surface brightness profile and a Gaussian line
profile with 300 km$\,$s$^{-1}$ FWHM (approximately twice the spectral
resolution) centered around
$\lambda_{\mathrm{Ly}\alpha}(z_\mathrm{QSO})$.  We emulated seeing
effects by convolving those nebulae with 2D Gaussians of the average
seeing FWHM of the particular observation.  Using the above defined
noise proxies we scaled the surface brightness of the nebulae by
integer multiples $n_\mathrm{spat}$ according to Eq.~(\ref{eq:4}).  We
downsampled our simulated nebulae to the grid of our datacubes and
added them before the final step of empirical PSF subtraction. By
visual inspection of the residual cubes we found that nebulae with
$n_\mathrm{spat}=5$ can be unambiguously discriminated from the
background. Exemplarily we show in Fig.~\ref{fig:detection_sim} the
results of this numerical experiment for nebulae covering the
$r_C=1.25$\arcsec{} aperture. Results for other aperture sizes were
similar, i.e. a 5$\overline{\sigma}^\mathrm{spat}$ input according to
Eq.~(\ref{eq:4}) yielded an unambiguous visual detection after PSF
subtraction.

We now demonstrate that both noise proxies yield similar results for
our surface brightness limits, defined as the minimum
surface brightness that a circular nebulae with a particular radius
could have before it would fall below our detection criterion.
For this purpose we show in Fig.~\ref{fig:detsigexample} the surface
brightness limits exemplarily for two objects, as a function of aperture radius.
These limits were calculated using either Eq.~(\ref{eq:4}) or Eq.~(\ref{eq:5}) with
$n_\mathrm{spat}=5$ or $n_\mathrm{spec}=5$. Figure~\ref{fig:detsigexample} shows 
the trend expected for a noise estimate independent of aperture, obtained by 
scaling the curve for the smallest aperture ($r_C=0.75$\arcsec{}) by 
aperture area $\propto N_{C_k}^{-1/2}$. Note that indeed
$\overline{\sigma}^\mathrm{spec}$ and
$\overline{\sigma}^\mathrm{spat}$ have similar values, and that there
is only a mild dependence on aperture size. For the cases not shown,
the agreement is similar. Since also no a priori distinction can
be made as to which of the two proxies is better, we quote the average
$(\overline{\sigma}^\mathrm{spec}+\overline{\sigma}^\mathrm{spat})/2$
as our detection limits in surface brightness.

\subsection{Surface brightness limits on extended emission}
\label{sec:surf-brightn-limits}

The resulting surface brightness upper limits for extended Ly$\alpha$ emission
surrounding the observed QSOs, calculated by applying the method
presented in the previous section, are shown in
Fig.~\ref{fig:radprofonepanel}. For reference, the obtained
upper limits for Ly$\alpha$ fuzz with 2.5\arcsec{} (20.5 kpc) radial
extend -- the typical extent of circum-QSO Ly$\alpha$ fuzz predicted by
\cite{Haiman2001} -- are (5.4, 3.4, 6.5, 6.0, 6.3) $\times$ $10^{-17}$
erg~s$^{-1}$ cm$^{-2}$ arcsec$^{-2}$ for the quasars Q0027$+$0103,
Q0256$-$0003, Q0308$+$0129, Q2243+0141, and UM\,247, respectively.

In Fig.~\ref{fig:radprofonepanel} we also show the integrated signal
with a circular aperture, i.e. Eq.~(\ref{eq:1}) transformed to surface
brightness units. Here the error bar on the integrated signal
indicates the standard deviation within the aperture. In our case,
this is a measure for how asymmetric possible signal is distributed
within the aperture; e.g. for UM\,247, a bright spot appears only to
the north of the object, thus the error bar on the integrated signal
within the circular aperture is large. Note that for all objects
except UM\,247 the integrated signal is always below our detection
limits, thus confirming the visual impression gained from
Figs.~\ref{fig:residual_images} and \ref{fig:residual_spec_plots}. We
discuss the fuzz around UM\,247 further in Sect.~\ref{sec:small-scale-lyalpha}.

Background offsets, positive or negative, can be seen in the
integrated signal for Q0308$+$0129 or Q2243$+$0141 - however, note
that we incorporated these systematics in the calculation of our
detection significances by forcing the expectation value to zero for
our noise proxies (see above).  There also appears a hint of possible
extended emission in Q0256$-$0003, which is however below our detection
threshold and not confidently separable from noise in the residual
datacube.

\subsection{Effect of sky subtraction on large-scale Ly$\alpha$ fuzz.}
\label{sec:effect-sky-subtr}

\begin{figure}
  \centering
  \includegraphics[width=0.5\textwidth]{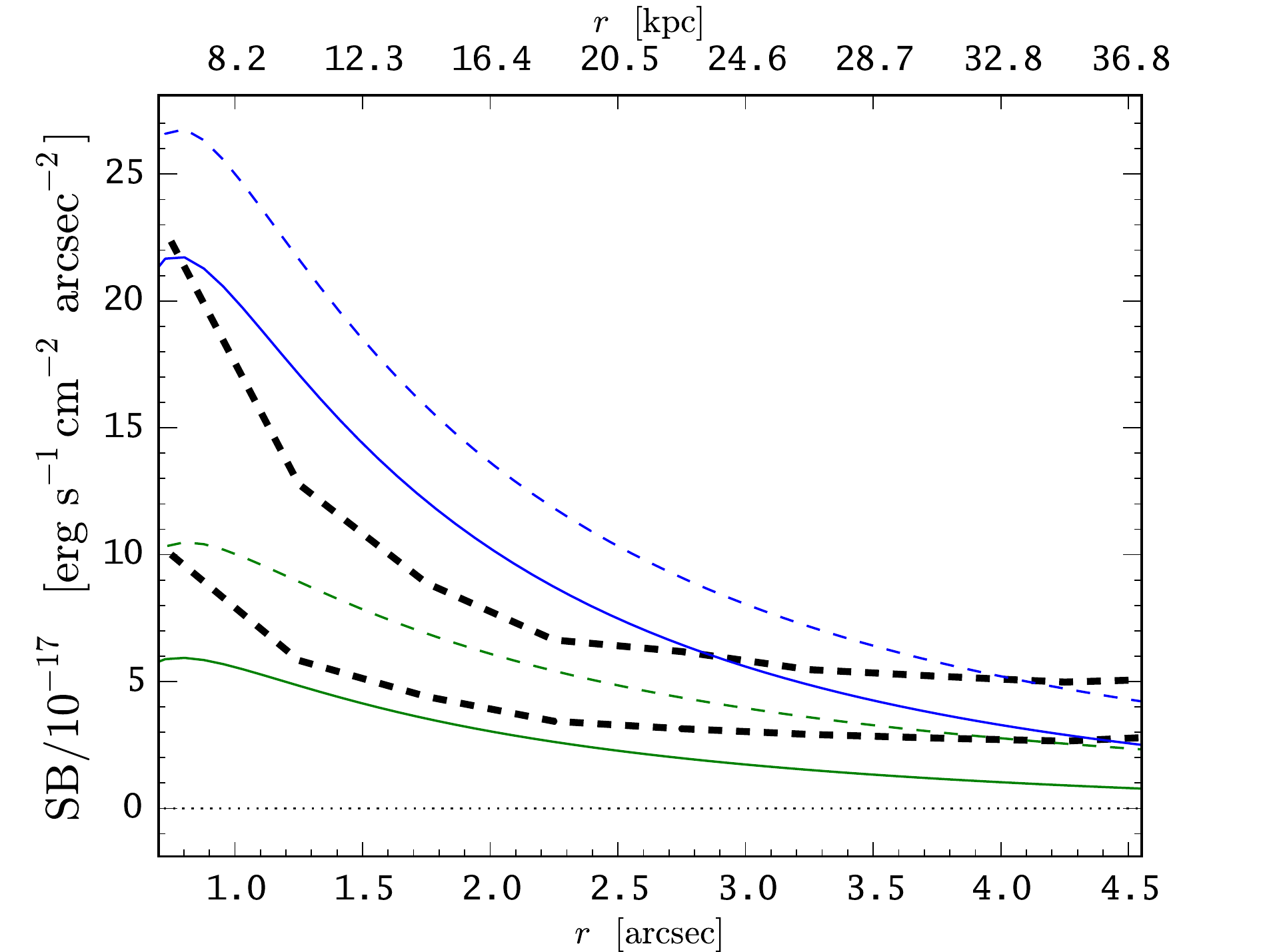}
  \caption{Expected average surface brightness profiles for exponential nebulae 
    $\mathrm{SB}_\mathrm{Ly\alpha}(r) = \Sigma_0
    \times \exp (-r/r_0)$, measured within circular apertures and 
    compared to the our surface brightness upper
    limits (\emph{dashed lines}, indicating the range of our limits
    shown in Fig.~\ref{fig:radprofonepanel}). Coloured \emph{solid lines} show the
    expectations including the effect of our sky subtraction
    procedure, while the \emph{dashed} lines ignore it (\emph{blue}:
    $\Sigma_0 \approx10^{-16}$erg\,s$^{-1}$\,cm$^{-2}$arcsec$^{-2}$
    and $r_0=2$\arcsec{}; \emph{green}: $\Sigma_0 = 5 \times
    10^{-17}$erg\,s$^{-1}$\,cm$^{-2}$arcsec$^{-2}$ and
    $r_0=4$\arcsec{}).}
  \label{fig:skysub_discussion}
\end{figure}

Recently, two nebulae were discovered that have sizes significantly 
larger than our field of view \citep[][see Table
\ref{tab:comp}]{Cantalupo2014,Martin2014}.  As explained in
Sect.~\ref{sec:sky-subtraction}, we subtracted a median spectrum
generated from the outer spaxels framing our field of view to
remove emission by the night sky.  Would giant nebulae as the above mentioned 
still leave a detectable signal in our observations?

From the intensity map presented in \cite{Cantalupo2014} we see that
at a distance of $r\approx5\arcsec{}$ to the QSO UM~287, 
their nebula has a mean surface brightness of
$\approx 5 \times 10^{-17}$~erg~s$^{-1}$ cm$^{-2}$ arcsec$^{-2}$. 
At a similar distance to their central QSO, \cite{Martin2014} measured
a surface brightness of $\approx 9 \times 10^{-18}$~erg~s$^{-1}$ cm$^{-2}$ arcsec$^{-2}$. 
If the surface brightness profile
increases towards the center, the central parts might still leave a
significant detectable signal in our datacubes.  If, however, the
central surface brightness profile is rather flat, the
significance of the recoverable signal will be substantially reduced.
Unfortunately, currently no information on the central parts of the
surface brightness profile in UM~287 is available (Cantalupo, priv. comm.).

We investigated the recoverability of the nebula in UM~287
assuming two different exponential surface brightness profiles,
$\mathrm{SB}_\mathrm{Ly\alpha}(r) = \Sigma_0 \times \exp (-r/r_0)$
with $r_0=2$\arcsec{} and $r_0=4$\arcsec{}, which are fixed to
UM\,287's surface brightness of
$10^{-17}$erg\,s$^{-1}$\,cm$^{-2}$arcsec$^{-2}$ at $r=5$\arcsec{}.
The central surface brightness of the $r_0=2$\arcsec{} profile is then
$\Sigma_0 \approx10^{-16}$erg\,s$^{-1}$\,cm$^{-2}$arcsec$^{-2}$ and
for the $r_0=4$\arcsec{} profile
$\Sigma_0 = 5 \times 10^{-17}$erg\,s$^{-1}$\,cm$^{-2}$arcsec$^{-2}$.
Our sky subtraction procedure would subtract a constant surface
brightness of $\mathrm{SB}(4\arcsec{})\approx 1.5 \times 10^{-17}$
from both profiles.  We note that such a faint signal would not have
been seen in the visual inspection of the subtracted sky spectra.

Applying our detection criterion in
Fig.~\ref{fig:skysub_discussion}, the $r_0=2$\arcsec{}
profile would still permit a significant detection of the nebula.
However, for the flatter $r_0=4$\arcsec{} profile the recovered
signal would fall below the detection threshold. And obviously we would 
always underestimate the true extent of the nebula.
We emphasize that our observations were originally not designed with
such large scale nebulae in mind and their recent discovery came as a
surprise.

\section{Discussion}
\label{sec:discussion--outlook}

\subsection{Comparison with observations from the literature}
\label{sec:comp-with-observ}

\begin{table*}
  \caption{Compilation of observational results from the literature on extended Ly$\alpha$ emission around radio-quiet QSOs (single-object investigations).}
\label{tab:comp}
  \centering  
  \begin{tabular}{lllcrl} \hline
    Reference & Quasar & \multicolumn{1}{c}{Redshift} & Extent & \multicolumn{1}{c}{Flux} & \multicolumn{1}{c}{Method}  \\ \hline
    \hline
    \cite{Bergeron1999} & Q\,2233-606 & 2.238 & 9.2\arcsec{}$\times$12.1\arcsec{} & 3.2$\times10^{-15}$ & narrow-band imaging \\
    \cite{Bunker2003}  & PC\,0953+4749 & 4.46 & $\approx$5\arcsec{} & 3.6$\times10^{-17}$ & long-slit spectrum \\
    \cite{Weidinger2005} & Q\,1205-30 & 3.041 & $\approx$10\arcsec{} & 7$\times10^{-16}$ &  narrow-band  \& long-slit \\
    \cite{Francis2006} & PSS\,2155+1358 & 4.28 & $\lesssim$0.7\arcsec{} & 1.7$\times10^{-17}$ & integral field spectroscopy \\
    \cite{Goto2012} & CFHQS\,J2329-0301 & 6.417 & $\sim$1.3 -1.5\arcsec{} & 8$\times10^{-17}$ & long-slit spectrum \\
    \cite{Rauch2013a} & Q\,J0332-2751 & 3.045 & 2.2\arcsec{} &  $ 5.4\times10^{-17}$ & long-slit spectrum \\
    \cite{Cantalupo2014} & UM\,287 & 2.28 & $\approx$55\arcsec{} & 5.5$\times10^{-15}$ & narrow-band imaging \\
    \cite{Martin2014}  & HS\,1449+19 & 2.843 &  $ \gtrsim$30\arcsec{} & 2.0$\times10^{-13}$ & integral-field spectroscopy \\ \hline \hline
  \end{tabular}  \tablefoot{Flux is given in erg\,s$^{-1}$cm$^{-2}$. Measurements on the fuzz
    around CFHQS\,J2329-0301 have also been published by
    \cite{Willott2011}, who obtained similar results. No flux has been
    reported by \cite{Bunker2003} for the  flux of the extended
    Ly$\alpha$ emission
    surrounding PC\,0953+4749, but this object is also part of the
    \citetalias{Christensen2006} sample. The flux for the nebula around HS\,1449+19 is from Table 3 in
    \cite{Martin2014}. See text for possible caveats
    considering the comparability of the presented values on flux and
    extent. }
\end{table*}

\begin{table*}
\caption{Compilation of observational results from the literature on
  extended Ly$\alpha$ emission around radio-quiet QSOs (samples).}\centering
\begin{tabular}{l@{}c@{\;\;}c@{}lll}\hline \hline
Reference & \multicolumn{2}{c}{Radio-Quiet Quasars} & Redshift & SB limit  & Instrument - Method \\
{}        & Targeted &  Fuzz detected   & {}       &       & 
 \\ \hline
\citetalias{Christensen2006} \textsuperscript{(a)} & 6 & 4 & 2.7 $\lesssim z \lesssim$ 4.5 &
                                                                       $\sim 10^{-17}$ & PMAS --  integral-field spectroscopy \\
\cite{North2012} \textsuperscript{(b)}      & 6 & 4 & $z\sim4.5$                    & $\sim5\times10^{-19}$ &  FORS2 - slit spectroscopy\\
\cite{Hennawi2013}     & 29 & 11 & $z\sim2$                    &
$\sim5\times10^{-18}$        &  LRIS \& GMOS - slit spectroscopy \\ \\
This study             & 5  & 1  & $z\sim2.3$                  & $\sim
                                                                 5.5\times10^{-17}$ & PMAS --  integral-field spectroscopy \\
\hline
\end{tabular}
\label{tab:comp_samp}
\tablefoot{(a)  Surface brightness limit for \citetalias{Christensen2006} estimated by us.
  (b) \cite{North2012} is an extension of
  \cite{Courbin2008}. Surface brightness limit in erg\,s$^{-1}$cm$^{-2}$arcsec$^{-2}$.}  
\end{table*}

\begin{figure}
  \includegraphics[width=0.5\textwidth]{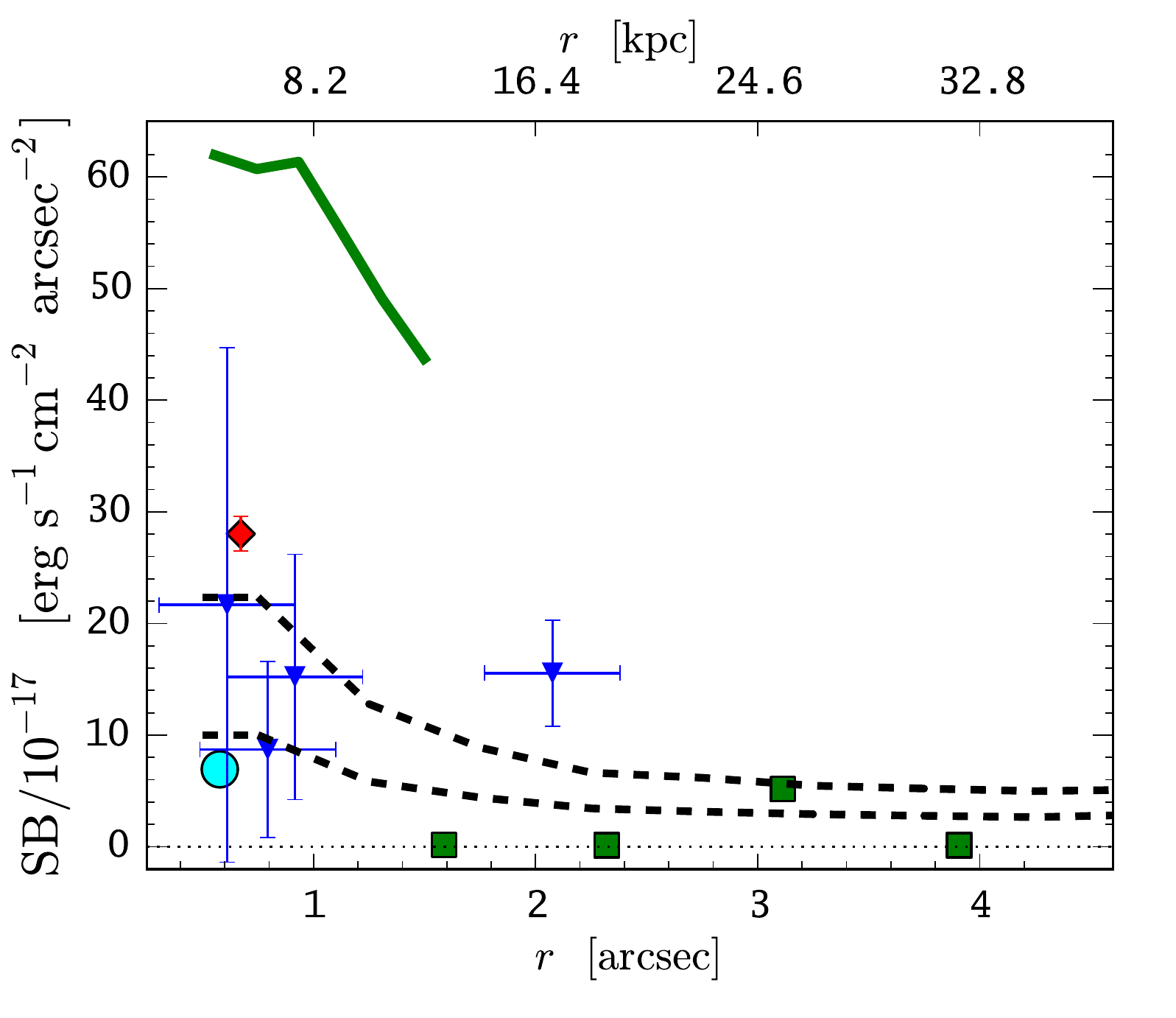}
  \caption{Comparison of our surface brightness upper limits as a
    function of radius (thick dashed lines) to the
    reported literature detections of circum-QSO
    Ly$\alpha$ emission, de-redshifted to $z=2.3$, and assuming
    that the reported maximum extent defines the radius of the
    detection aperture. The symbols feature results from the
    \citetalias{Christensen2006} sample (\emph{blue triangles} with
    error bars), from the \cite{North2012} sample (\emph{green
      squares}), the nebula from \cite{Goto2012} (\emph{red diamond})
    and the fuzz from \cite{Francis2006} (\emph{cyan circle}). We also
    show the integrated signal at various radii from the de-redshifted
    \cite{Weidinger2005} surface brightness profile (\emph{solid green
      line}).}
  \label{fig:comp_all}
\end{figure}

How do our upper limits in surface brightness for extended Ly$\alpha$
emission around radio-quiet QSOs compare with previous investigations
of this phenomenon?  Most of the studies reporting a successful
attempt in detecting circum-QSO Ly$\alpha$ fuzz (sometimes serendipitous
discoveries) have focused on single objects. We compiled a
list of such investigations in Table~\ref{tab:comp}. Beyond these
single-object results, very few studies aimed at constructing actual
samples (which were always small). We list the relevant publications in
Table~\ref{tab:comp_samp}.

Substantial methodological differences between the studies listed in
Tables~\ref{tab:comp}~\&~\ref{tab:comp_samp} have to be kept in mind
when comparing those results with our non-detections.  The sizes in
Table~\ref{tab:comp} often refer to the maximum extent at which the
authors were able to detect Ly$\alpha$ emission.  This quantity obviously
depends on the depth of the observations and, in the case of long-slit
spectroscopy, the orientation of the slit. For the latter case, fluxes
are also affected by significant slit-losses, since only a fraction of 
the nebula is usually captured. Finally, line emission from the central QSO
might contaminate the fluxes of the nebular component in some cases, 
especially since a subtraction of the QSO point source was not performed 
in some cases \citep[e.g.][]{Bergeron1999,Martin2014,Rauch2013a}. 

The quoted values for the surface brightness limits of the samples in
Table~\ref{tab:comp_samp} are also very rough estimates, as the
actual limits depends on the assumed size of the fuzz
\citep[see our derivation in Sect.~\ref{sec:calc-detect-limits} and
also the derivation in Sect.~4.3 of ][]{Hennawi2013} and, moreover,
they differ from target to target due to differences 
in the used instruments and observing strategies. 
Unfortunately, \citetalias{Christensen2006} did not
quantify the depth of their observations. Since they also used PMAS
(although in a different setup than we) and knowing the typical 
instrumental and atmospheric parameters, we estimated that
their detection sensitivity was similar to that in our study.

We next considered whether the Ly$\alpha$ fuzz detected by the studies 
listed in Tables~\ref{tab:comp} and \ref{tab:comp_samp} would have been
recovered if `implanted' into our quasars.  For this exercise we
selected only objects for which the reported maximum
extent (after de-redshifting) would be covered by our field of view.  
The \cite{Martin2014}, \cite{Cantalupo2014} and \cite{Bergeron1999} 
objects do not fulfill this criterion (see also Sect.~\ref{sec:effect-sky-subtr}). 
We also excluded the \cite{Rauch2013a} object, since their flux value is
contaminated by quasar Ly$\alpha$ emission. We then circularized the
nebulae, i.e.\ we assumed the flux to be distributed symmetrically around
the quasar with a radius defined as half the maximum extent. After
calculating de-redshifted radii and surface brightness levels,
we determined the integrated signals within circular apertures
covering the whole nebulae. Note that only \cite{Weidinger2005} 
provided a surface brightness profile, so that only for this object
we could calculate the integrated signal at various radii.

In Fig.~\ref{fig:comp_all} we plot the results from this exercise and
compare them to our surface brightness limits: Out of 11 nebula used in
this calculation, 3 would be detected in all our
observations, further 4 would yield a 5$\sigma$ detection only in our deepest
dataset, and 4 would not be recoverable at all. (Note however that the
circularization is actually reducing the signal from sources that have
significantly asymmetric flux distributions. Thus a nebula yielding a
5$\sigma$ detection only in our deepest dataset might have been
recovered at a higher significance if the area over which the signal
was integrated was better matched to the light distrubution of the nebula.)

Note that half of the points in Fig~\ref{fig:comp_all} are at radii
less than 10 kpc and 8 out of 10 are below 25 kpc.  Small scale
Ly$\alpha$ fuzz appears to be quiet frequent around radio-quiet QSOs.
This is also supported by the high recovery rate of this phenomenon in
\cite{Hennawi2013}, although their detections are generally
fainter. We argue in the next section that the signal we find around
UM\,247 also falls into this category.

Finally, we point out that most of the reported detections of bright
Ly$\alpha$ fuzz are typically at $z\gtrsim 3$.  Our recovery rate of
this phenomenon at $z\approx2.3$ (20\%) is lower compared to
\citetalias{Christensen2006} (66\%). Moreover, \cite{Hennawi2013}
report a recovery rate of 38\% at $z\sim2$. Observations are
insufficient at this stage to infer a redshift evolution of circum-QSO
Ly$\alpha$ fuzz properties. It is nevertheless intriguing that
\cite{Zirm2009} find that Ly$\alpha$ nebulae enshrouding radio-loud
QSOs decline in size and luminosity with decreasing redshift. If real,
such a decline could indicate a depletion of the cool gas content of
the CGM (at least on average). Extending such a trend to radio-quiet
QSOs would however require much larger and more homogenous samples
over a range of redshifts.

\subsection{Small-scale Ly$\alpha$ fuzz near UM\,247}
\label{sec:small-scale-lyalpha}

UM\,247 is the only object in our sample with a formally significant residual 
at Ly$\alpha$ after PSF subtraction. To assess whether this residual is real 
and not an artefact caused by a fault (e.g. an
undetected cosmic) in a single exposure, we repeated the stacking of the 
individual exposures (Sect.~\ref{sec:stacking}) 5 times, each time
with one exposure excluded from the stack. We then performed the
empirical PSF subtraction in exactly the same manner as described
above (Sect.~\ref{sec:psf}) for each of those stacks. The feature
remained, hence we assert that it is genuine.

This basically unresolved excess emission is located at a distance of 
$d\lesssim$4-8 kpc (0.5-1\arcsec{}) north of UM\,247. The line has a
flux of $\sim 5.6 \times 10^{-16}$ erg\,s$^{-1}$cm$^{-2}$,
corresponding to a Ly$\alpha$ luminosity of $L_\mathrm{Ly\alpha}
\approx 2.4 \times 10^{43}$ erg\,s$^{-1}$.

If the Ly$\alpha$ emission of this feature was powered
purely by star formation, ignoring ionization of the nearby quasar
as well as radiative transfer effects, this would correspond to a
star formation rate of $\sim 10$--20~$\mathrm{M}_\odot\,\mathrm{yr}^{-1}$ 
\citep[with the exact value being dependent on the metallicity of the underlying stellar
population, see][]{Schaerer2003}.  This Ly$\alpha$ luminosity
is the bright end of the Ly$\alpha$ emitter luminosity
function for $z\sim2.3$ \citep{Blanc2011}. Given the small field of view
of our observations, the detection of such a bright Ly$\alpha$ emitter
in close vicinity to a quasar would appear rather coincidental. 

In an opposing scenario we consider the possibility that the Ly$\alpha$ 
radiation is produced purely by fluorescence of a compact nearby cloud, 
devoid of internal star formation and optically thick to ionizing radiation 
from the quasar. This cloud will behave as a special mirror, converting
$\sim$66\% of all impinging hydrogen-ionizing photons into Ly$\alpha$.
Thus, given the quasar ionizing luminosity and the size of the cloud
one can predict its Ly$\alpha$ luminosity 
\citep[e.g.][]{Francis2004,Adelberger2006,Kollmeier2010, Hennawi2013}.
Since the cloud is unresolved, we adopt as an upper limit of its size
the extent of one PMAS spaxel.  Thus the physical surface extent of
the Ly$\alpha$ emitting cloud is $R_\mathrm{cloud}^2 \lesssim$ 16
kpc$^2$. From a measured QSO flux density of $f_\lambda = 2.5 \times
10^{-16}$\,erg\,s$^{-1}$cm$^{-2}$\AA{}$^{-1}$ for UM\,247 at 4500\AA{},
we estimate a quasar luminosity at the Lyman edge of
$L_{\nu_{\mathrm{LL}}} = 8.0 \times 10^{29}$\,erg\,s$^{-1}$Hz$^{-1}$,
assuming a power-law index of
$\alpha_1 = 0.44$ ($f_\nu\propto\nu^{-\alpha}$) 
for the quasar continuum redward of the Lyman edge
\citep{VandenBerk2001}, and $\alpha_2 = 1.57$ blueward of $912$\AA{} \citep{Telfer2002}. 
We also corrected for galactic extinction
($A_\lambda=0.08$ at 4500\AA{} for UM\,247), although that is a small effect. 
Assuming isotropic radiation, this results in an ionizing photon number
flux of $\Phi = 4\times10^{10}$~s$^{-1}$~cm$^{-2}$
($1\times10^{10}$~s$^{-1}$~cm$^{-2}$) at a distance of
$d=4$\,kpc (8\,kpc). The expected Ly$\alpha$ luminosity of a spherical cloud 
at this location follows as
\begin{equation}
  \label{eq:8}
  L_\mathrm{Ly\alpha} = 4 \pi \times f_\mathrm{gm} \times
  \eta_\mathrm{thick} \times h\nu_\mathrm{Ly\alpha} \times
  R_\mathrm{cloud}^2 \times \Phi \,\mathrm{,}
\end{equation}
where $\eta_\mathrm{thick}$ is the fraction of ionizing continuum
photons converted to Ly$\alpha$ photons,
i.e. $\eta_\mathrm{thick}=0.66$.  With $f_\mathrm{gm}$ we denote the
geometric reduction factor -- a free parameter that accounts for the
inhomogeneous illumination of the cloud and subsequent redistribution
of Ly$\alpha$ photons over a wide solid angle. Radiative transfer
simulations suggest that for a transversely illuminated cloud
$f_\mathrm{gm} = 0.5$ \citep{Kollmeier2010}.  For our upper limit on
the cloud radius, Equation (\ref{eq:8}) then provides an upper limit
on its Ly$\alpha$ luminosity of $L_\mathrm{Ly\alpha}
\lesssim 4 \times 10^{44}$~erg~s$^{-1}$ 
($1 \times 10^{44}$~erg~s$^{-1}$) at $d=4$\,kpc (8\,kpc). This is
almost an order of magnitude higher then the observed value. In
reality, however, the cloud size might be much smaller than our
instrumentally imposed upper limit, and it might also be further away
than the projected transverse distance. Hence this order of magnitude
estimate of Ly$\alpha$ radiation emanating from the surface of an
optically thick cloud in the vicinity of the quasar is still consistent
with what we observe.

\subsection{Comparison with models}
\label{sec:lyalpha-fuzz-generic}

\cite{Haiman2001} presented a strongly idealized model to predict the
luminosity of Ly$\alpha$ fuzz around radio-quiet quasars.
Specifically, they assumed a spherical symmetric 2-phase gas
distribution in pressure equilibrium within a collapsed dark-matter
halo and predicted extended Ly$\alpha$ emission as a generic property of
high-$z$ QSOs. The phases are a hot tenuous virialized plasma
(i.e. $T_\mathrm{hot} \sim T_\mathrm{vir}$ of the halo) and colder
neutral gas that has cooled down to $T_\mathrm{cold}\sim10^4$~K during
the age of the system. If in such an environment the quasar 
emits radiation isotropically and ionizes the whole nebula, 
its Ly$\alpha$ luminosity depends only on the total gas mass
and thus on the total halo mass. In this framework, the absence
of significant Ly$\alpha$ fuzz in our QSOs might suggest that the
hosting haloes are not overly massive. However, as already pointed out by
\cite{Haiman2001} and also \cite{Alam2002}, small deviations from this
idealized scenario might alter the surface brightness of the
Ly$\alpha$ fuzz substantially. Currently, the observations do
not provide strong constraints on the basic assumptions in these
models. 

State-of-the-art numerical simulations in a cosmological framework
predict that the spatial distribution of the CGM gas shows filamentary
structure, with cold gas accreting along streams towards the center of
the halo \citep[e.g.][]{Dekel2009,Faucher-Giguere2011,Rosdahl2012}.
Those streams have high column densities and their surfaces will
therefore reflect up to 2/3 of incoming ionizing photons as Ly$\alpha$
photons. Although Ly$\alpha$ cooling radiation from those streams
alone might already produce a detectable signal in extremely massive
haloes, the presence of a central quasar should enhance the
contrast of the filamentary structures by boosting their
Ly$\alpha$ emissivity by up to 2 orders of magnitude \citep{Kollmeier2010}.
While the giant Ly$\alpha$ nebulae around some quasars may well 
be explained by such fluorescently glowing accretion streams, these are
by no means typical. More high-resolution simulations will be required 
in order to predict robustly the luminosities and sizes of Ly$\alpha$ fuzz 
for the lower-mass haloes typical for radio-quiet QSOs.

\section{Conclusions}
\label{sec:summ-conl-remarks}

It seems that on average, radio-quiet QSOs are rather unspectacular
sources of spatially extended Ly$\alpha$ emission. The nondetection of 
such Ly$\alpha$ fuzz in 4 of our objects and the marginal detection in
one case are all fully consistent with the results of other recent investigations, 
although even the combined samples are still small. 

The spectacularly bright and extended Ly$\alpha$ nebulae discovered 
around a few quasars \citep{Weidinger2005,Martin2014,Cantalupo2014} 
must be probably considered very rare cases. The rarity of this phenomenon 
may be explained if giant Ly$\alpha$ nebulae are seen only around QSOs 
that reside in extraordinarily massive haloes. 

The observing techniques used to search for Ly$\alpha$ fuzz around QSOs
are quite divers, encompassing narrow-band imaging, long-slit spectroscopy,
and now also integral field spectrocopy (IFS). In principle, IFS should
surpass the other methods by a large margin; in fact, any IFS datacube
allows the user to explore both the narrow-band imaging as well as the
spectral domain. The limiting factor for most existing optical IFS studies
-- including the present investigation --
is sensitivity and light-collecting area of the available telescopes. 
This is however about to change, as a new generation of
efficient IFS systems is being deployed at 8--10m class telescopes.
Of particular interest for the topic of this study is the MUSE instrument,
recently commissioned at the ESO Very Large Telescope \citep{Bacon2014}. 
Its unprecedented sensitivity will make it an optimal discovery machine for
Ly$\alpha$ fuzz around quasars. Moreover, the large field of view of MUSE
will ensure that there is no longer a danger of sky subtraction removing
physical signal from very extended nebulae with flat radial profiles. It is
to be expected that within a relatively short time, the statistics of
observed Ly$\alpha$ fuzz around quasars will improve dramatically, turning
the emphasis from discovery to the detailed dissection of physical properties.

\begin{acknowledgements}
  We thank the support staff at Calar Alto observatory for help with
  the visitor-mode observations.  E.C.H. especially thanks Sebastian
  Kamann and Bernd Husemann for teaching him how to operate the PMAS
  instrument. For all visual inspections of IFS datacubes mentioned in
  this paper we used the software
  \texttt{QFitsview}\footnote{\url{http://www.mpe.mpg.de/~ott/QFitsView/}}
  by \cite{Ott2012}. All plots were made with \texttt{matplotlib}
  \citep{Hunter2007}.  The color scheme in
  Fig.~\ref{fig:residual_images} is the \texttt{cubehelix} color
  scheme by \cite{Green2011}.  We thank Sebastiano Cantalupo for
  sharing unpublished details about the nebula around UM\,287 with
  us. Finally, we thank the anonymous referee for constructive input.
\end{acknowledgements}

\end{document}